\documentclass{aa}

\usepackage{graphicx}
\usepackage{amsmath}
\usepackage{longtable}
\usepackage{txfonts}
\usepackage{natbib}
\usepackage{multirow}
\usepackage{layout}
\usepackage{url}
\usepackage{nicefrac}
\usepackage{keyval}
\usepackage{wasysym}
\usepackage[colorlinks=false,pdfborder={0 0 0}]{hyperref}
\bibpunct{(}{)}{;}{a}{}{,}

\begin{document}

   \title{Age and metallicity gradients in fossil  ellipticals}
   \titlerunning {Age and metallicity gradients in fossil  ellipticals}

   \author{P.\ Eigenthaler\inst{\ref{wien},}\inst{\ref{puc}} \and W.\ W.\ Zeilinger\inst{\ref{wien}}}
   \institute{
   Institut f\"ur Astrophysik, Universit\"at Wien, T\"urkenschanzstra\ss e 17, A-1180 Vienna\label{wien}
   \and
   Departamento de Astronom\'{i}a y Astrof\'{i}sica, Pontificia Universidad Cat\'{o}lica de Chile, Av.\ Vicu\~{n}a Mackenna 4860, Santiago, Chile \email{eigenth@astro.puc.cl}\label{puc}
   }

   \date{Received xxx; accepted xxx}

   \abstract
   {Fossil galaxy groups are speculated to be old  and highly evolved  systems of galaxies  that formed  early in the universe and had enough time  to
   deplete their $L^{*}$     galaxies through successive mergers of  member galaxies, building  up one massive  central elliptical, but  retaining the
   group  X-ray  halo.}
   {Considering that  fossils   are    the remnants    of mergers    in    ordinary groups, the     merger  history of    the    progenitor group   is
   expected    to  be  imprinted in      the fossil  central  galaxy (FCG).  We  present for  the first
   time radial  gradients of single-stellar  population  (SSP)   ages  and  metallicites in a sample   of FCGs to constrain their  formation
   scenario. We also measure   line-strength   gradients for the strongest  absorption    features in these galaxies.}
   {We took deep  spectra with   the  long-slit spectrograph ISIS  at   the William  Herschel Telescope (WHT) for six FCGs.    The obtained spectra
   are fit    with    Pegase$\,$HR  SSP models  within  the  full-spectrum fitting  package ULySS  yielding SSP  ages and metallicities of the stellar
   populations. We  measure    radial gradients  of SSP ages and metallicities along the major axes. Lick indices are measured for the
   strongest absorption  features    to determine line-strength gradients and compare with the full-spectrum fitting results.}
   {Our  sample   comprises some   of  the  most  massive   galaxies in   the  universe  exhibiting   an   average central   velocity  dispersion   of
   $\sigma_0=271\pm28$   km s$^{-1}$.  Metallicity gradients  are throughout  negative with  comparatively flat   slopes  of  $\nabla_{[\rm{Fe/H}]}=-
   0.19\pm0.08$  while age   gradients are   found  to  be  insignificant  ($\nabla_{\rm{age}}=0.00\pm0.05$).       All  FCGs  lie on the  fundamental
   plane,  suggesting that  they  are virialised systems.  We  find  that   gradient strengths and central metallicities are  similar  to  those  found 
   in cluster    ellipticals of similar mass.}
  {The  comparatively   flat   metallicity gradients with  respect  to those predicted by  monolithic  collapse ($\nabla_{Z}=-0.5$)  suggest
  that  fossils  are indeed the result of  multiple major mergers.  Hence we  conclude that   fossils are not 'failed groups' that formed 
    with a top heavy luminosity function. The low scatter  of gradient slopes suggests  a   similar merging history for  all galaxies   in our sample.}

   \maketitle


\section{Introduction}
Fossil    galaxy    groups    are    observationally    defined    as       extended     X-ray     sources    with      X-ray    luminosities    above
$L_{X,\,\textrm{\scriptsize bol}} = 0.5 \times 10^{42}$      $h_{50}^{-2}$   erg~s$^{-1}$ and   a  single bright elliptical     dominating    the  optical,   all
other galaxies within  one-half  virial   radius being at  least    2  magnitudes  fainter   in  the  $R$ band (\citealt{jones03}). Earlier  numerical
simulations  of  \citet{barnes89} predicted such   observational  characteristics for  old and evolved   groups  of galaxies  that had enough time  to
deplete their $L^{*}$     galaxies  through successive mergers of  member  galaxies, building  up one massive  central elliptical, but  retaining  the
group-sized  X-ray   halo. Contrary to  this hierarchical merging  scenario it  was also suggested  that fossils  could simply be 'failed groups'  that
formed with a top-heavy luminosity function absent of $L^{*}$ galaxies \citep{mulchaeyzabludoff}.

The    first   fossil    group        was   identified   by \citet{ponman94}.   Since then,   several   other                     objects
have                   been assigned to                this                              class               in                the                last
few                  years   (\citealt{vikhlinin,jones03,yoshioka,sun04,khosroshahi04,ulmer,schirmer10,pierini11,miller12}).           \citet{jones03}
estimated  fossils   to be quite  frequent,   constituting   8-20\%     of     all   systems   with       comparable     X-ray   luminosity      ($\ge
10^{42}$   erg~s$^{-1}$). The detection of  these aggregates  is  challenging   however, mainly  because  of  their unremarkable  appearance  in   the
optical  and  the   lack of high $S/N$   X-ray data.  Recent   studies   have     shown   that  fossils      are much       more massive          than
expected         for remnants        of         group-sized         haloes,        comparable         to          poor       clusters           rather
than         groups (\citealt{mendesdeoliveira06,mendesdeoliveira09,cypriano,proctor11,eigenthaler12}).    Because  of  these  inconsistent  findings,
mainly  due to  the small number  of  fossils  studied so far, various    formation scenarios  have  been  proposed  for   these systems.   Based  on
scaling  relations  of  ordinary  groups  and  clusters,    \citet{khosroshahi07} concluded    that  fossils   had   an       early  formation  epoch,
while \citet{proctor11} stated   that  fossils   might simply    be   dark  clusters, being   comparable  to  ordinary  clusters  in   mass    and  
X-ray luminosity,     but otherwise  possessing    the  richness and   optical  luminosities  of    poor  groups.  Simulations   by    \citet{dariush07}
using the     Millennium  and     Millennium    gas   simulations  (\citealt{springel05}) revealed   that  at  any    given  redshift, fossils    have
assembled  a  higher fraction  of    their  final   halo    mass     compared  to  non-fossils   with  similar  X-ray luminosity,   suggesting    that
fossils    indeed    formed    early  with most   of  their    halo   mass   in  place    at  higher   redshifts. A   similar   result   was  found by
\citet{donghia05}  using  N-body/hydrodynamical  simulations,    concluding that  fossils     have already  assembled  half     of  their  final  dark
matter      mass    at    $z    \apprge     1$,  leaving  sufficient      time   for   $L^{*}$       galaxies   to      merge,   resulting    in   the
exceptional magnitude gap    at $z=0$  whilst  in    non-fossils the  final dark matter   halo    mass is    assembled quite recently. Although fossil
systems are believed to have collapsed early and have assembled most of their virial masses at higher redshifts, the central galaxies in fossil groups
may have merged later than non-fossil bright central galaxies \citep{diazgimenez08}.

Assuming that fossils originate  from ordinary galaxy groups that  evolved to compact groups which subsequently   coalesced into  one massive elliptical, the
merger  history of the   progenitor group    is expected to  be   imprinted in     the fossil  central galaxy   (FCG) and     any differences   in the
evolution      compared    to field or cluster ellipticals  should  be reflected   in  the   overall  characteristics of       these  galaxies.  Since
the space density  of fossils      is    considered to be    as  large    as  that  of     poor    and   rich    clusters combined,   fossils    could
also   provide   suitable    environments   for   the   formation   of   luminous cD  galaxies   observed   in  clusters. Following   this    possible
connection,   \citet{khosroshahi06b}  investigated  the  morphology  of  seven FCGs    with  the    conclusion  that     the isophotal      shapes  of
FCGs      are    always     discy       in     contrast     to  brightest  group (BGG)   and   cluster    (BCG)   galaxies. \citet{labarbera09}  found  no differences in  the    structural
parameters   between  a  sample  of  25    FCGS, selected   from   the  SDSS,  and   field ellipticals.  The      examination of stellar   populations
revealed   similar   ages, metallicities,  and  $\alpha$-enhancement compared   to   bright field  ellipticals.   In    addition, elliptical  galaxies
in compact groups    were found to  be older  and   more metal  poor  than field ellipticals and fossil groups, which had already  been  established in  earlier   works
(\citealt{proctor04,delarosa07}).   It  was  thus   noted that   FCGs  cannot  be formed  by  dry  mergers  of ellipticals    in
compact   groups.  However,   a  wet    merger   with   a   gas-rich    disk  galaxy   could   also   explain  the observed relations. More  recently,
\citet{mendezabreu12}  have  studied  the   near-infrared photometric    and   structural     properties   of   a sample     of     20  FCGs.    These
fossils   closely    follow    the  observed   tilted    fundamental    plane    of     normal   ellipticals (\citealt{bernardi03}), suggesting   that
they are     dynamically relaxed    systems that    suffered dissipational    mergers  during      their formation  (\citealt{hopkinsfp}). However,  a
change   in  the    slope  of  scaling      relations involving  stellar    mass   was  interpreted    by  subsequent  major  dissipationless mergers,
increasing  size  and mass   of   the  galaxies but   leaving  velocity dispersion unaltered   (\citealt{bernardi11b}). Because of low  S/N data these
authors couldn't  determine  whether fossils  are  systematically boxy  or  discy.  Since  most    of these  previous  studies have   focused  on  the
photometric and   structural  properties of FCGs,  detailed information   on their spatially resolved  stellar  population parameters is still lacking.

This paper   investigates age,   metallicity  and   line-strength gradients    of a    sample  of   six FCGs    to    shed light on  the formation and
evolution  of  fossils.  The  galaxies  were       selected    from   the     samples     of    \citet{mendesdeoliveira06},     \citet{santos},    and
\citet{eigenthaler09},    according to their   redshift     and    high  apparent brightness.  At   redshifts  between   $0.023 \lesssim   z  \lesssim
0.081$,    the   systems   lie     at luminosity distances  between    $100   \lesssim D_{L}  \lesssim    370$ Mpc. These   values  correspond to   an
age  of the  universe   of  13.1  back to   12.4  Gyr at  the redshift  of the   observed galaxies.  According to  \citet{dariush07}, the   Millennium
simulation predicts   that  fossils  at these   redshifts   should  have  assembled at    least   95\%  of   their   total dark  matter     mass. 
 
 The
paper   is organised  as     follows.  In Sect.\    2     observations    and  data   analysis    techniques      are   described  whilst    Sect.\  3
presents   all  obtained   results.  Sect.\   4 discusses   and   compares   our findings  with   similar data    from     the  literature.
Conclusions   are  given   in Sect.\  5.    Magnitudes presented  in      this work are        SDSSIII   DR9  model      magnitudes. Throughout    the
paper,    the  standard  $\Lambda$CDM   cosmology with $\Omega_{M}=0.3$, $\Omega_{\Lambda}=0.7$, and  a  Hubble constant of  $H_{0}=70$ km~s$^{-1}$ is
used.

\section{Observations and data reduction}
Long-slit spectroscopy  of a sample  of six  FCGs was performed  with the    long-slit spectrograph ISIS   at the William Herschel Telescope  (WHT) in
visitor mode. The   observations were carried   out  in  2008   on December 20   and in 2009   on April 28.     ISIS     is suitable     for long-slit
spectroscopy  up     to    4     arcmin slit    length.    Through  the   use  of   dichroic    filters,      the    instrument  allows   simultaneous
observations  in both    a    blue  and       a red     arm.  Both   arms    provide    an  array    of  4096$\times$2048 pixels  with 13.5$\mu$m  and
15$\mu$m  pixel  size, respectively.    The  CCDs are arranged so  that    the  long axis   is  placed along     the dispersion   direction maximising
spectral   coverage.   The spatial scale    is  0.2 arcsec  pixel$^{-1}$  in   the  blue and   0.22  arcsec    pixel$^{-1}$  in  the red  yielding   a
maximum  unvignetted  slit-length  of  3.7 arcmin.   The spectrograph  is  designed such that    a  slit-width  matching  the typical seeing   of  the
site  ($\sim$1 arcsec)   corresponds   to  approximately 4   pixels   FWHM   on  the    CCD  for  all   gratings.  To   cover  the  most   prominent
absorption   features  H$\beta$,      Mg$_2$, Mg$b$,    Fe5270,    Fe5335, Fe5406,  the      6100  dichroic  in  combination  with the  R300B  grating
(0.86 \AA\space  pixel$^{- 1}$) in the   blue and   the    R1200R grating  (0.26   \AA\space    pixel$^{-1}$)  in  the   red   was     requested.   If
no   emission  lines     were    expected around  H$\alpha$  from   SDSS  spectra,   the  R316R    grating  (0.93   \AA\space   pixel$^{    -1}$)  was
used   instead.   A slit  width     of    1 arcsec was  considered    for all galaxies.

Table  \ref{ellipticals}     lists   all     ellipticals        observed     at   the     WHT.  The    total integration  time  of the   targets   was
split   into 3  or  4   single exposures    of  $\sim45$    minutes each       to     correct for cosmic   rays.  The spectrograph slit was aligned to
the  major axis   of  each galaxy  except for RX  J0752.7+4557 and RX  J0844.9+4258  where the  slit was aligned  so  that nearby bright galaxies were
also  covered. The slit position  angles are given in Table \ref{ellipticals}.

\begin{table*}
\begin{minipage}[t]{520pt}
\caption{\label{ellipticals}Fossil ellipticals observed with ISIS.}
\centering
\renewcommand{\footnoterule}{}
\begin            {tabular*}{\textwidth}{@{\extracolsep{\fill}}p{2.5cm}p{1.5cm}p{1.6cm}p{0.7cm}p{1.1cm}p{1.4cm}p{0.5cm}p{2cm}p{2cm}p{2cm}}
\hline
\hline
 \multicolumn{1}{c}{\multirow{2}{*}{galaxy}}         &  \multicolumn{1}{c}{\multirow{2}{*}{$\alpha_{2000}$}}   &  \multicolumn{1}{c}{\multirow{2}{*}{$\delta_{2000}$}}   & \multicolumn{1}{c}{\multirow{2}{*}{$z$}} &  \multicolumn{1}{c}{$r'$}                    &       \multicolumn{1}{c}{exposure}         &    \multicolumn{1}{c}{PA}                   &                 \multicolumn{2}{c}{spectral range}                     &  \multicolumn{1}{c}{FWHM}          \\
                                                     &                                                         &                                                         &                                          &          \multicolumn{1}{c}{[mag]}           &          \centering{[seconds]}             &           \multicolumn{1}{c}{[$^{\circ}$]}  &          \centering{blue arm}     &    \centering{red arm}             &   \multicolumn{1}{c}{[arcsec]}     \\
\hline
    NGC 1132                                         &    \centering{02  52  51.8}                             &     \centering{$-$01  16  28.8}                         &   \centering{0.023}                      &  \hspace{5pt}{12.19}                   &       \centering{4 $\times$ 2700}          &      \multicolumn{1}{c}{140}                &   \centering{3800$-$6250$\,$\AA}  &   \centering{6600$-$7200$\,$\AA}   &   \multicolumn{1}{c}{1.3}          \\
    RX J0752.7+4557$^{a}$                            &    \centering{07  52  44.2}                             &     \centering{  +45  56  57.4}                         &   \centering{0.052}                      &  \hspace{5pt}{14.43}                   &       \centering{3 $\times$ 3000}          &      \multicolumn{1}{c}{ 76}                &   \centering{3800$-$6250$\,$\AA}  &   \centering{6600$-$7200$\,$\AA}   &   \multicolumn{1}{c}{1.3}          \\
    RX J0844.9+4258$^{a}$                            &    \centering{08  44  56.6}                             &     \centering{  +42  58  35.7}                         &   \centering{0.054}                      &  \hspace{5pt}{13.93}                   &       \centering{3 $\times$ 3000}          &      \multicolumn{1}{c}{ 36}                &   \centering{3800$-$6250$\,$\AA}  &   \centering{6600$-$7200$\,$\AA}   &   \multicolumn{1}{c}{1.3}          \\
    RX J1152.6+0328$^{b}$                            &    \centering{11  52  37.6}                             &     \centering{  +03  28  21.8}                         &   \centering{0.081}                      &  \hspace{5pt}{14.37}                   &       \centering{3 $\times$ 2700}          &      \multicolumn{1}{c}{ 23}                &   \centering{3800$-$6250$\,$\AA}  &   \centering{6300$-$8500$\,$\AA}   &   \multicolumn{1}{c}{0.7}          \\
    RX J1520.9+4840$^{b}$                            &    \centering{15  20  52.3}                             &     \centering{  +48  39  38.6}                         &   \centering{0.074}                      &  \hspace{5pt}{13.68}                   &       \centering{3 $\times$ 2600}          &      \multicolumn{1}{c}{ 94}                &   \centering{3800$-$6250$\,$\AA}  &   \centering{6300$-$8500$\,$\AA}   &   \multicolumn{1}{c}{0.7}          \\
    RX J1548.9+0851$^{a}$                            &    \centering{15  48  55.9}                             &     \centering{  +08  50  44.4}                         &   \centering{0.072}                      &  \hspace{5pt}{13.20}                   &       \centering{3 $\times$ 2600}          &      \multicolumn{1}{c}{ 33}                &   \centering{3800$-$6250$\,$\AA}  &   \centering{6300$-$8500$\,$\AA}   &   \multicolumn{1}{c}{0.7}          \\
\hline
\end              {tabular*}
\begin{footnotesize}
\begin{flushleft}
{\bf Notes:} \\
$^{a}$ from  \citet{santos}                                  \\
$^{b}$ from  \citet{eigenthaler09}                           \\
\end{flushleft}
\end{footnotesize}
\end{minipage}
\end              {table*}

\begin{figure*}
\centering
\includegraphics[width=\textwidth]{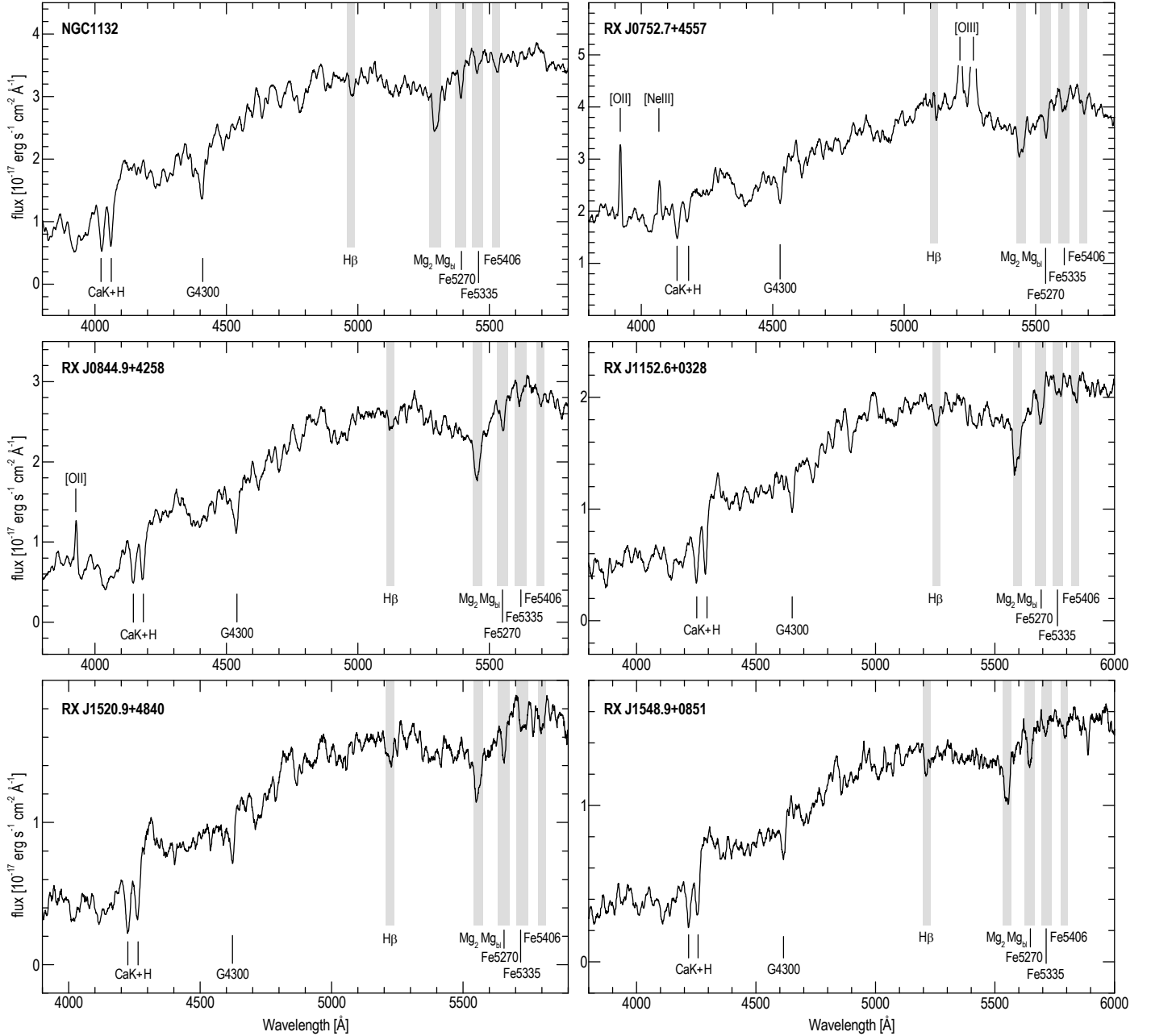}
\caption{\label{spectra}  Observed  spectra  in   the  ISIS  blue  arm  smoothed to   the Lick/IDS  resolution   ($\sim  9\,$\AA). The most prominent
spectral features and Lick/IDS index passbands (grey) are highlighted. The spectra corerspond to the central pixel of each galaxy.}
\end{figure*}

Basic  data reduction and calibration was carried out in IRAF  based on  bias,  flat, and arc frames taken in  the afternoon and at dawn.    Arcs were
also  taken during the night after each movement of the telescope to account  for possible detector shifts.   General  CCD  reduction  steps   such as
bias    subtraction,    trimming,    and flat-fielding   were    carried  out  within    the        IRAF {\tt    ccdred}  package    to   remove   the
instrument    signature. Bad  pixels were  removed by   interpolating over nearby  pixel regions. Subsequent spectral  reduction was performed in  the
{\tt longslit}  package. Small geometric distortions  introduced by  the camera  optics were removed by  tracing the  spectrum   on  the detector  and
transforming  the tilted    coordinates  to the orthogonal image  axes.  Spectra were  wavelength   calibrated by   identifying    ISIS  CuNe and CuAr
lines  in   the arc  lamp spectra.  Line  positions  were fit   by   a    3$^{\rm{rd}}$    order   cubic spline     to  set  the dispersion  solution.
Individual  frames  were average   combined  to  improve   the   S/N and   filter   out  cosmic   rays.  Sky subtraction  was  carried out  with   the
IRAF {\tt background}  task by  specifying  two   or  three  background windows  on  either  side  of  the combined  galaxy  spectra.  A $6^{\rm{th}}$
order Chebyshev  polynomial was      then fit  to  these    background regions   in all   columns  along    the dispersion   direction and  subtracted
from  the data.  The fit to the sky was checked for all  galaxies in steps of 100 pixels along the dispersion direction. We note that  because of
the small  spatial extent of  the investigated galaxies with  respect  to the CCD chip, the   Chebyshev polynomial was always  flat  or  showed only a
very minor linear  trend along the galaxy which was always much  smaller  than the noise of the  sky background.  The background windows  were chosen so that a   large
area of the sky  was covered along the  slit. The comparatively  high degree of the Chebyshev polynomial  was  chosen so that a good  estimate for
the sky was achieved along the   whole slit.  The  polynomial was  almost always  found to  be flat.  The sky   subtracted spectra   were  then  checked
individually for  any remaining  night sky emission  lines. For  the  fainter emission  lines, the subtraction  yielded an   excellent  result.   Only
the     strongest     lines,   [OI]  at  $\lambda           5577$\AA \space        and       the  sodium  doublet   at   $\lambda   5890$\AA    \space
showed   minor   residual  patterns.      ING       spectrophotometric     standard         stars  matching  the       airmass          of         the
scientific                targets                 were              selected               from                  the              ING        landscape
tool\footnote{\url{http://catserver.ing.iac.es/landscape/}}. Standard  star    spectra   were  extracted     by  means   of  the  algorithm  described
in  \citet{horne86} to    retrieve  the  maximum S/N   without  biasing   the resulting    fluxes.  To  minimise  light   loss   due   to differential
atmospheric refraction,  all stars    were  observed  with  the    slit oriented  along   parallactic    angle. Flux   calibration  was performed   by
comparing   the observed   spectrophotometric  standard  star   count  rates    with   the  corresponding   absolute   flux   tables    in   La Palma
Technical    Notes  65   and  100.  To  account    for      atmospheric   extinction,   the   data    from     La   Palma    Technical   Note  31  was
applied.    Due  to   the long exposure  times,    effective airmasses   have    been    calculated     for    each
individual   galaxy via    the formula      from  \citet{stetson89}. The  resulting   sensitivity    functions were   finally  applied    to   the
galaxy spectra,  removing both atmospheric extinction and  the sensitivity  characteristics  of  the  CCD.   To  transform    observed    
 absorption-line features   into  the  Lick IDS    system,      Lick    standard      stars    were     selected    from      the       homepage      of      Dr.\
Guy   Worthey\footnote{\url{ http://astro.wsu.edu/worthey/html/system.html}}.   To match   the   stellar  populations   of  the  observed galaxies,   G
and  K giants were   considered for  that purpose.   Figure \ref{spectra}  shows  the final reduced  spectra  of  the central pixel  of each galaxy in
the  blue arm smoothed  to the resolution of the Lick/IDS spectrograph.

\section{Data analysis and results}
A  reliable determination  of   ages   and   metallicities  in    ellipticals  is difficult,  since   both  quantities alter   spectra
and colours   of  these  galaxies in    a similar  way.   An    age change    of     $+30$\%  is    cancelled     by  a  simultaneous      metallicity
change     of        $-20$\%.     This      is      widely      known       as      the  age-metallicity       degeneracy,     or   $3/2$   degeneracy
(\citealt{worthey99}).      Individual       spectral        features         sensitive        to      either     age       or metallicity     can  be
used    to    break    this       degeneracy,   however    (\citealt{burstein84,worthey94,worthey97}).     These    well-calibrated  features    allow
predictions  for  the  absorption-line     strengths  of     whole single stellar     populations   (SSPs) of     a    given age       and metallicity
(\citealt{worthey94}).    When  focusing  only    on specific  features  in    the  spectrum    of  an      old  stellar   population,  one  obviously
misses  the wealth   of information  embedded    in  all  the    remaining pixels    along    the  dispersion direction,   however.    To account  for
this   additional   information  besides   the    strongest  absorption  features,  full   spectrum  fitting  techniques   have   been      developed.
Because of the very low  S/N  in  absorption lines  at larger radii of  the observed ellipticals,  ages and  metallicities   are  solely based
on  full-spectrum fitting  in this work.

\subsection{Full  spectrum  fitting  with ULySS} 
 The                   open-source            package      ULySS\footnote{\url{http://ulyss.univ-lyon1.fr/}}   (University          of            Lyon
 Spectroscopic  Analysis         Software;   \citealt{koleva09})     was  used to      fit    synthetic   single        stellar    population    (SSP)
 models   directly   to      the   observed     galaxy  spectra    yielding SSP equivalent  ages  and   metallicities   of  the   stellar  populations
 of  the   investigated   galaxies.      ULySS   makes      use of    the Pegase$\,$HR    models      from        \citet{leborgne04}  based         on
 the    Elodie$\,$3.1     library      of   stellar             spectra (\citealt{elodie3}).  Since    the    synthesis    of    a stellar  population
 requires   a  stellar   spectrum   at    any  point    in       the  parameter    space  ($T_{\rm   eff}$,    log~$g$    and       [Fe/H]),       the
 library      involves  an   interpolator     for  that   purpose.     All  Pegase$\,$HR   models     are  computed    assuming   a   Salpeter     IMF
 and   Padova  1994    evolutionary      tracks   providing synthetic  SSPs with   ages   between   1$ -$20000$\,$Myr     and  metallicities   between
 $-2.3$  and $0.69\,$dex.    A multiplicative  polynomial  is   used  to  adjust the  overall  spectral  shape  to   the   SSP    model.    The  order
 of the     polynomial can    be  freely    chosen  but      shouldn't   be   taken too     low resulting   mostly in    a   severe mismatch of    the
 investigated  spectrum  and   the  SSP     model. \citet{koleva09} have shown that  the model  fits converge to stable values between  $10\le n \le
 15$ for the spectra of  F, G, K and O  stars and $n > 35$  for an M star and  becoming unstable again above $n  = 150$. Hence, in  order  to  achieve
 reasonable, stable    values    an   order   of 40   was chosen except where   otherwise stated.

\begin{figure*}
\centering
\includegraphics[width=350pt]{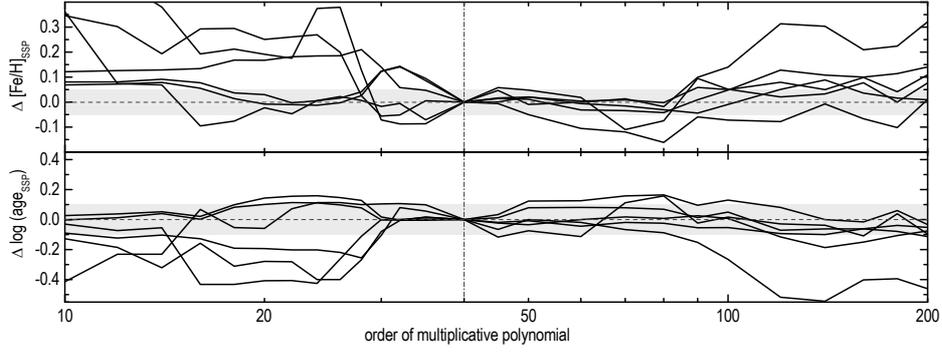}
\caption{\label{nplot}  Variation  of SSP  age  and  metallicity    with  increasing  order  $n$  of   the  multiplicative
polynomial for all observed ellipticals.   The deviations to  values at $n=40$ are displayed. Deviations corresponding to 0.05 dex in [Fe/H] and 0.1 dex in age are shaded.}
\end{figure*}

\begin{figure}
\centering
\includegraphics[width=\columnwidth]{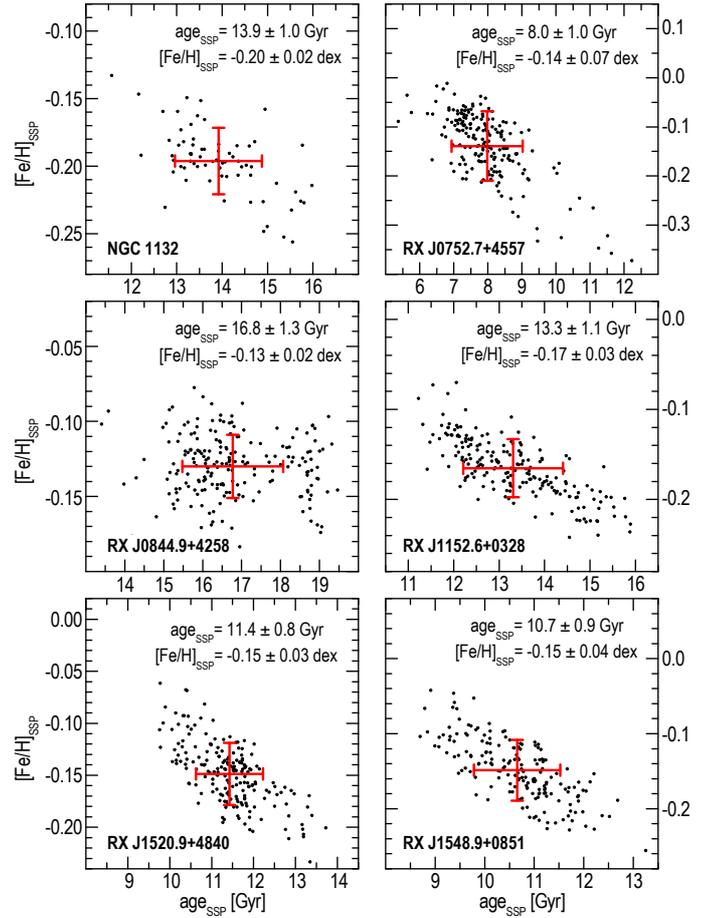}
\caption{\label{agemetallicity}ULySS   fit results for   the integrated stellar   populations of all   observed FCGs. The   point distributions result
from  200 Monte-Carlo  simulations.  Adopted final values and  errors for  age and metallicity are the averages and  standard  deviations of the shown
point distributions indicated by the bold crosses. Obvious outliers have been excluded for the measurements.}
\end{figure}

 Figure \ref{nplot} presents   the variation of   the  fit results   with  increasing order  of    $n$, specifically   the   deviations to values    at
$n=40$. Shaded    areas   correspond  to  deviations of   0.05  dex in [Fe/H] and 0.1  dex in age. Except for  RX J1548.9+0851,    values stay  within
0.05  dex  in  [Fe/H]  from $40$$\,$$\le$$\,$$n$$\,$$\le$$\,$$80$,  matching  the  average  error    of the  measured values. For orders above $n=80$,
values start to diverge again since the polynomial is trying to over-fit the spectra. Values of age show a larger spread  compared to metallicity. The
variations of age and metallicity with increasing $n$ are likely due to the limited S/N of the observed spectra. The ULySS  parameter  {\tt       \textbackslash CLEAN}   was considered   for  the fitting to    exclude possible  outliers  in the galaxy   spectrum,
resulting  from  any remaining minor night sky  emission. The clipping algorithm reduces the sensitivity of a fit to such undesirable features by  reiterating   the   fit   after     outliers    in     the   residuals   have   been    identified   and   masked\footnote{ see   \url{http://ulyss.univ-
lyon1.fr/uly.html\#ULY_FIT}  for  a detailed  decription.}.   The  reliability    of   the     performed SSP   fits was tested through the use of  $\chi^{2}$    and
convergence  maps in   the    age-metallicity  parameter    space.   To  determine   integrated SSP  equivalent ages   and
metallicities    of   each   galaxy, one-dimensional spectra were   created   by  summing  up all
the signal  along  the spectrograph slit.  200  Monte-Carlo simulations   were computed   for every galaxy,   repeating a    fit   successively   with
random Gaussian noise\footnote{The    dimension  of   the   added   noise was  based    on a   user-defined  signal-to-noise  ratio   provided     for
each  galaxy.    The S/N     ratios  were      measured  with    IRAF {\tt      splot} at      a rest-frame      wavelength      of  $\sim5140\,$\AA.}
applied   to   the spectrum   in       each   step.    The resulting      point    distributions     were     then   used    to   calculate    average
SSP    ages   and  metallicities   and     the   associated   standard  deviations.     Outliers   clearly     detached     from   the  main     point
distributions     were excluded for     these   measurements. Only   blue  arm  spectra were   considered   because of the  lack  of   prominent
absorption   features   in  the red channel.   Convergence maps  confirmed     the  independence  of    the obtained results   from  varying   initial
guesses for  age and  metallicity. Figure \ref{agemetallicity}   shows    the  final  point distributions of age and  metallicity resulting from  the
applied Monte-Carlo simulations   of the  one-dimensional  SSP   fits.   While   the  galaxies  spread    from    8 to    17 Gyr  in   SSP model  age,
SSP   metallicities  show   consistent subsolar   values around     -0.15 dex.   Almost all   Monte-Carlo simulations    follow a   similar   trend,
revealing a  much higher  uncertainty in  age than in metallicity. 

\subsection{Age- and metallicity  gradients}
\begin{figure*} \centering \includegraphics[width=420pt]{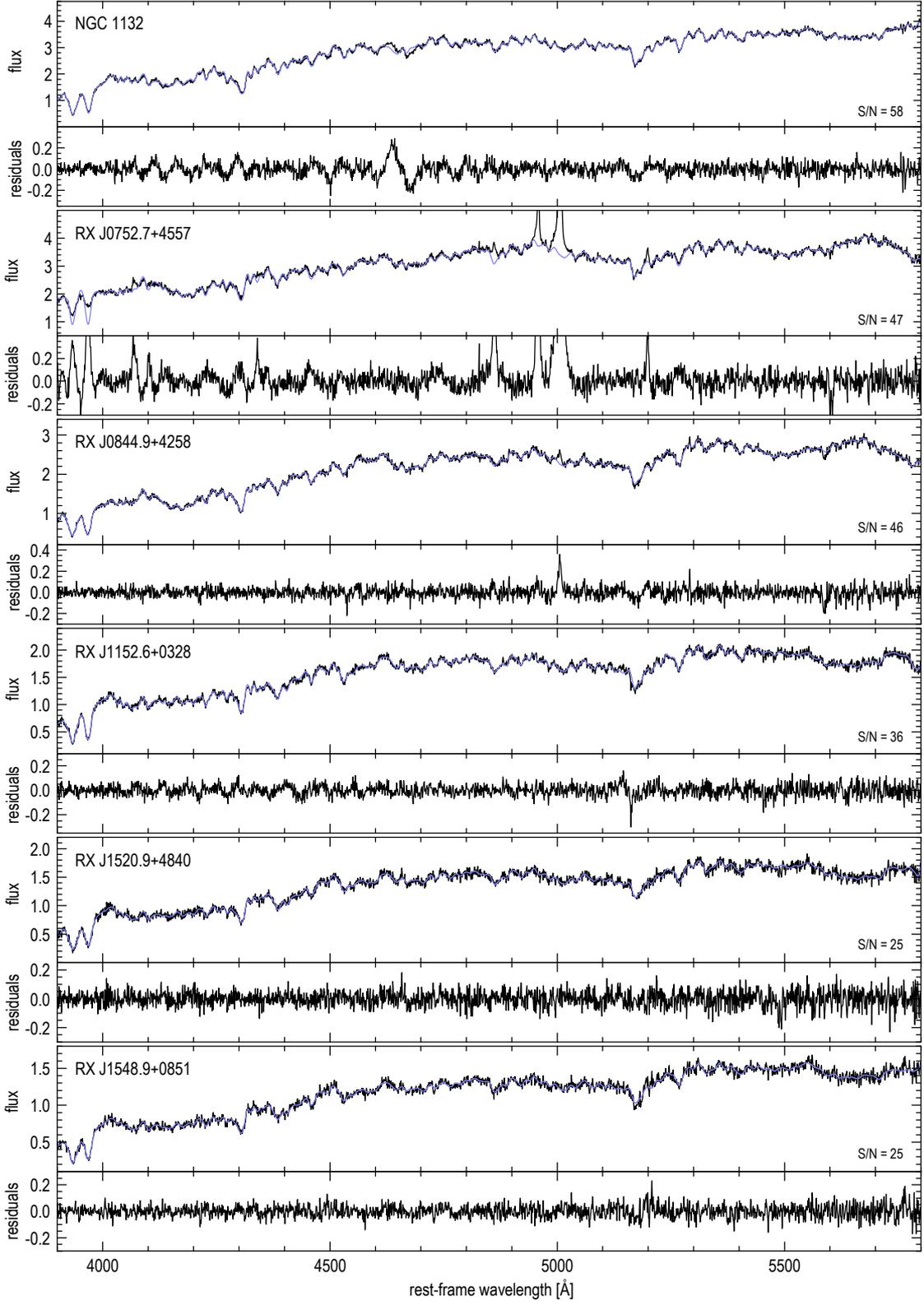} \caption{\label{ulysscentral}  ULySS SSP fits to the central bin of the  observed
FCGs. The galaxy  spectra are shown  in black  while  the best-fit SSP  models are shown  in blue. Fluxes  are given in 10$^{-17}$ erg s$^{-1}$
cm$^{-2}$ \AA$^{-1}$. S/N ratios of the investigated spectra are also presented. } \end{figure*}

\begin{figure*}
\centering
\includegraphics[width=420pt]{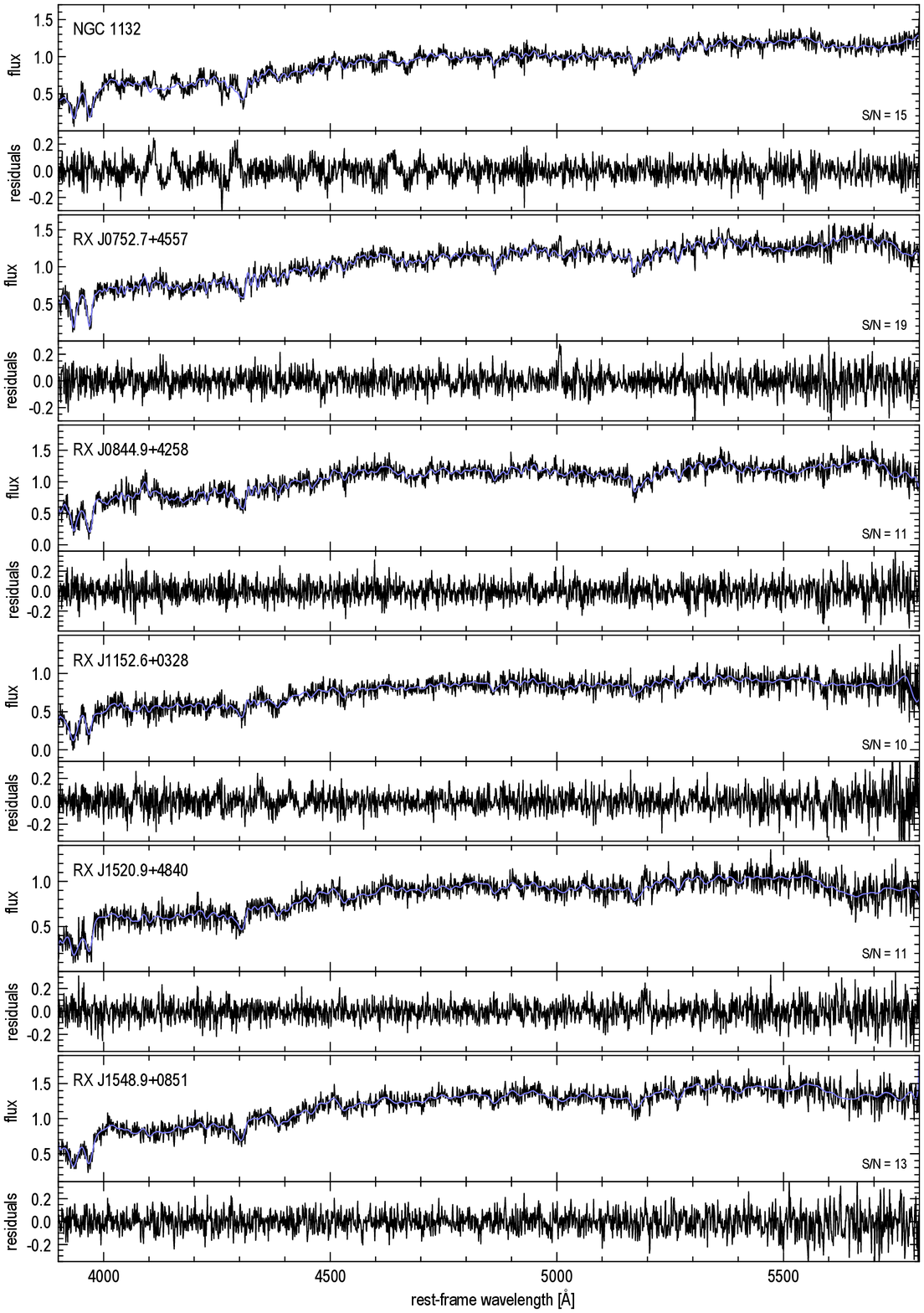}
\caption{\label{ulyssout}  Same as Fig.\ \ref{ulysscentral} but for the outermost bins. Fluxes  are given in 10$^{-18}$ erg s$^{-1}$ cm$^{-2}$ \AA$^{-1}$.}
\end{figure*}

\begin{figure*}
\centering
\includegraphics[width=\textwidth]{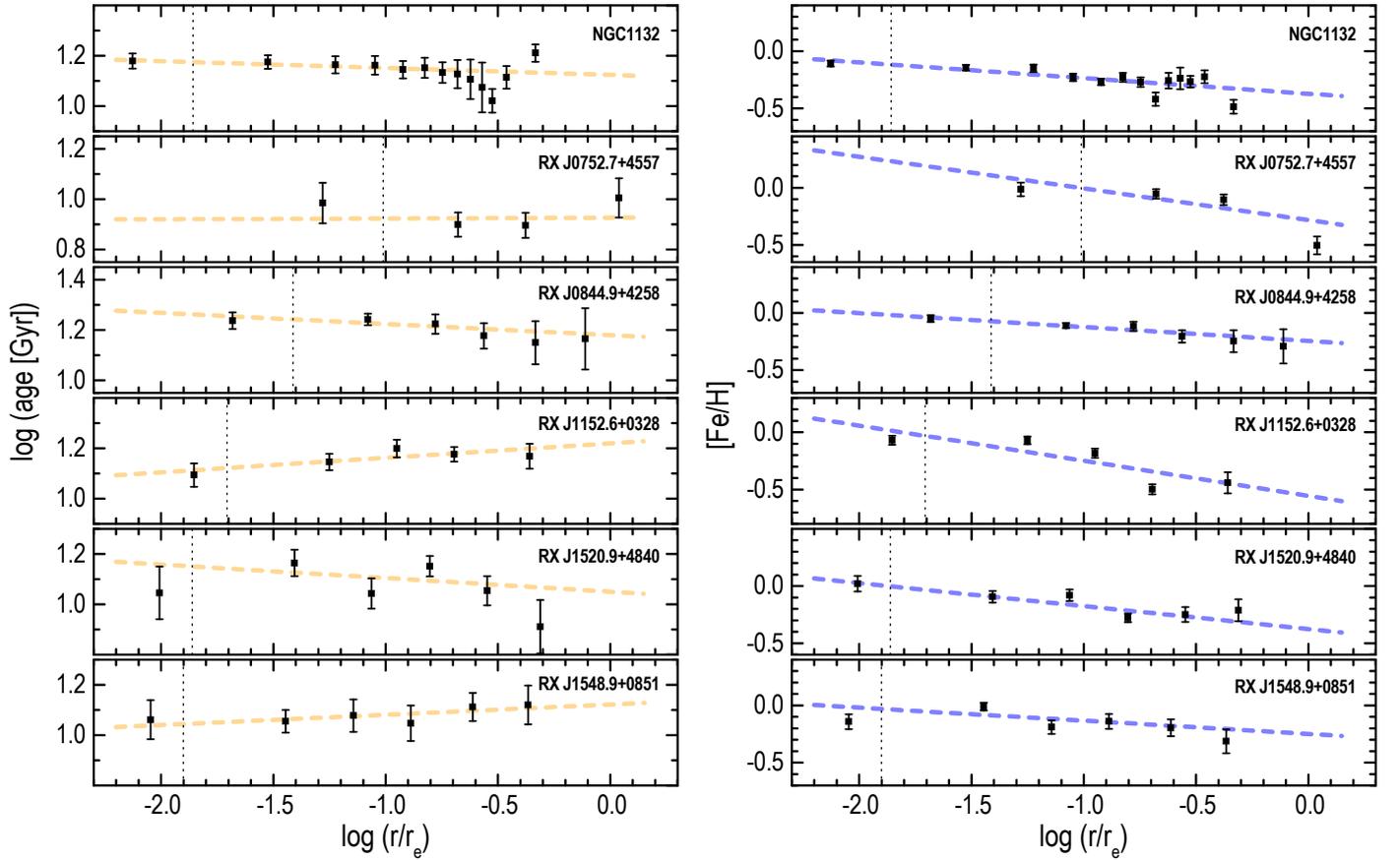}
\caption{\label{gradients}Radial  gradients  of  age  and  metallicity  for  the  observed  FCGs.  Left  panels  refer  to  age,  right  ones  present
metallicity. Dashed  lines show the linear relations fit to the observed data.  The ordinates are the decimal logarithm of age and the metallicity  in
dex.  Each  datapoint is  the result   of 200  Monte Carlo   simulations and  was averaged  from both  sides of  the galaxy centre. Error bars are the
corresponding standard deviations. Metallicity  gradients are  throughout negative  ($-0.31\le  \nabla_{[\rm{Fe/H}]}\le -0.11$)  while  age  gradients
are   flat scattering  around zero  ($-0.05\le \nabla_{\rm{age}}\le 0.06$). Vertical lines indicate the seeing-dominated central parts.}
\end{figure*}

To investigate  gradients   of   age  and  metallicity along     the  galaxy   major  axis,   the observed long-slit spectra    were    chopped   into
bins    along  the   spatial    axis.   The  individual    spectra were   then     fit  separately       with ULySS. Each  spectrum was binned   along
the spatial  direction  to achieve    a   higher S/N.  The seeing     was estimated    to 1.3   and 0.7 arcsec for   the  two  nights    corresponding
to  about        7  and    4  pixels,   respectively.   Central galaxy spectra were  binned only by this small number of pixels, matching the  present
seeing while  $30-40$ pixel rows were binned in the outermost spectra to      achieve  a  reasonable S/N along    the whole  slit.   Again, the  S/N was  estimated   for each   spectrum
with  {\tt          splot}  at   a rest-frame   wavelength       of $\sim5140\,$\AA. The  S/N varies from $\sim 60-25$ for the central bins to   $\sim
19-10$ for  the outermost  bins. Figures  \ref{ulysscentral} and  \ref{ulyssout} show   spectra of  the central  and outermost  bins together with the
corresponding SSP model fits and  residuals.  The   estimated    noise  strongly   affects    the   resulting Monte-Carlo
simulations. While   the  central    bins showed  well-defined  point    distributions,  the   outer   ones   yielded  much more   diffuse    results.
In addition,  there  is  a  clear tendency that  the  fits in the    outer  bins  tend to    converge  to   the  model   limit  at  19.99 Gyr.   These
data   points were  all  neglected   for   the age  and  metallicity measurements.  In  each  bin, the  resulting  ages and metallicities   were  then
computed as   the    mean  of   the      200  Monte Carlo       simulations.   At   a  given  radius,   ages  and  metallicites  were  averaged  from  both  sides  of  the  galaxy  centre.
The  corresponding error  bars  were estimated  from the associated
standard deviations. Assuming power-law
gradients  in age  and metallicity,    linear relations   between $\log\rm{age}$,  [Fe/H], and  $\log  r/r_{\rm{eff}}$  are  expected.  Hence, linear
relations of the form

\begin{align}
\log \left( {{\rm{age }}[r/{r_{{\rm{eff}}}}]} \right) &= \log \left( {{\rm{ag}}{{\rm{e}}_{{\rm{eff}}}}} \right) + {\nabla _{{\rm{age}}}}\log \left( {r/{r_{{\rm{eff}}}}} \right) \label{gradientage}\\
\left[ {{\rm{Fe/H}}} \right][r/{r_{{\rm{eff}}}}] &= {\left[ {{\rm{Fe/H}}} \right]_{{\rm{eff}}}} + {\nabla _{\left[ {{\rm{Fe/H}}} \right]}}\log \left( {r/{r_{{\rm{eff}}}}} \right) \label{gradientmetallicity}
\end{align}

have been fit to the  SSP ages and metallicities obtained  with ULySS,  where $\nabla_{\rm{age}}$  and  $\nabla_{\left[{{\rm{Fe/H}}}\right]}$  are the
corresponding  gradients  in age   and  metallicity.   The   fits    are   presented     in    Fig.\   \ref{gradients}    while    the   corresponding
numbers  and  input   parameters  for  the   fits  are   given    in  Table \ref{ellipticalsresults}.  

 All   metallicity gradients  exhibit  negative
slopes   between   $-0.11\le  \nabla_{\left[{{\rm{Fe/H}}}\right]}    \le  -0.31$   while    gradients  in age   show   both   positive   and  negative
slopes  between  $-0.05\le \nabla_{\left[{{\rm{Fe/H}}}\right]}  \le 0.06$.   Gradient errors were determined directly from  the linear least   squares
fits.  To quantify  the goodness of  the fits,  the coefficient of
determination  $R^{2}$ is listed in Table  \ref{ellipticalsresults}   for all fits. A value   of $R^{2}=1$ corresponds to a  perfect   linear relation
while $R^{2}=0$  indicates no linear relation at  all.  The observed metallicity profiles are much better fit by the  utilized power-laws
than the profiles in age.

\begin{table*}
\begin{minipage}[t]{520pt}
\caption{\label{ellipticalsresults}SSP   equivalent  central  ages  and   metallicites, central   velocity  dispersions  and  gradients   in age   and
metallicity obtained  with  ULySS. The corresponding fit parameters are also given.}
\centering
\renewcommand{\footnoterule}{}
\begin            {tabular*}{\textwidth}{@{\extracolsep{\fill}}p{2.5cm}p{0.3cm}p{1.8cm}p{1.3cm}p{1.6cm}p{0.6cm}p{0.3cm}p{0.3cm}p{1cm}p{0.3cm}}
\hline
\hline
 \multicolumn{1}{c}{galaxy}         &      \multicolumn{1}{c}{\tt MD{$^{a}$}}   &  \multicolumn{1}{c}{\tt LMIN$-$\tt LMAX$^{b}$}   & \multicolumn{1}{c}{age$_{0}\,$[Gyr]}    &  \multicolumn{1}{c}{[Fe/H]$_{0}$}           &       \multicolumn{1}{c}{$\sigma_{0}$ [km s$^{-1}$]$^{c}$}            & \multicolumn{1}{c}{$\nabla_{[\rm{Fe/H}]}$$^{d}$}           &  \multicolumn{1}{c}{$R^{2\,e}$}       & \multicolumn{1}{c}{$\nabla_{\rm{age}}$$^{d}$}          &  \multicolumn{1}{c}{$R^{2\,e}$}            \\
\hline
NGC 1132                            &                \centering{50}             &   \centering{3800$-$6100$\,$\AA}                 &  \multicolumn{1}{r}{15.1$\,\pm\,$1.0}   &  \multicolumn{1}{r}{-0.11$\,\pm\,$0.03}     &      \multicolumn{1}{l}{\hspace{0.2cm}  227$\,\pm\,$ \space 3}  &   \multicolumn{1}{c}{\hspace{-0.3cm}$-0.14\,\pm\,0.02$}    &   \multicolumn{1}{c}{0.68}            & \multicolumn{1}{r}{\hspace{-0.2cm}$-0.03\,\pm\,0.02$}  &   \multicolumn{1}{c}{0.13}                 \\
RX J0752.7+4557                     &                \centering{40}             &   \centering{3800$-$5800$\,$\AA}                 &  \multicolumn{1}{r}{ 9.7$\,\pm\,$1.8}   &  \multicolumn{1}{r}{-0.01$\,\pm\,$0.06}     &      \multicolumn{1}{l}{\hspace{0.2cm}  296$\,\pm\,$18       }  &   \multicolumn{1}{c}{\hspace{-0.3cm}$-0.28\,\pm\,0.17$}    &   \multicolumn{1}{c}{0.57}            & \multicolumn{1}{r}{\hspace{-0.2cm}$ 0.00\,\pm\,0.08$}  &   \multicolumn{1}{c}{0.00}                 \\
RX J0844.9+4258                     &                \centering{60}             &   \centering{3800$-$5800$\,$\AA}                 &  \multicolumn{1}{r}{17.3$\,\pm\,$1.3}   &  \multicolumn{1}{r}{-0.05$\,\pm\,$0.03}     &      \multicolumn{1}{l}{\hspace{0.2cm}  284$\,\pm\,$ \space 4}  &   \multicolumn{1}{c}{\hspace{-0.3cm}$-0.12\,\pm\,0.02$}    &   \multicolumn{1}{c}{0.87}            & \multicolumn{1}{r}{\hspace{-0.2cm}$-0.04\,\pm\,0.02$}  &   \multicolumn{1}{c}{0.48}                 \\
RX J1152.6+0328                     &                \centering{40}             &   \centering{3800$-$5800$\,$\AA}                 &  \multicolumn{1}{r}{12.4$\,\pm\,$1.3}   &  \multicolumn{1}{r}{-0.07$\,\pm\,$0.04}     &      \multicolumn{1}{l}{\hspace{0.2cm}  246$\,\pm\,$ \space 5}  &   \multicolumn{1}{c}{\hspace{-0.3cm}$-0.31\,\pm\,0.14$}    &   \multicolumn{1}{c}{0.62}            & \multicolumn{1}{r}{\hspace{-0.2cm}$ 0.06\,\pm\,0.03$}  &   \multicolumn{1}{c}{0.59}                 \\
RX J1520.9+4840                     &                \centering{40}             &   \centering{3800$-$5800$\,$\AA}                 &  \multicolumn{1}{r}{11.1$\,\pm\,$2.7}   &  \multicolumn{1}{r}{ 0.02$\,\pm\,$0.07}     &      \multicolumn{1}{l}{\hspace{0.2cm}  279$\,\pm\,$ \space 9}  &   \multicolumn{1}{c}{\hspace{-0.3cm}$-0.20\,\pm\,0.06$}    &   \multicolumn{1}{c}{0.71}            & \multicolumn{1}{r}{\hspace{-0.2cm}$-0.05\,\pm\,0.08$}  &   \multicolumn{1}{c}{0.10}                 \\
RX J1548.9+0851                     &                \centering{40}             &   \centering{3800$-$5800$\,$\AA}                 &  \multicolumn{1}{r}{11.5$\,\pm\,$2.1}   &  \multicolumn{1}{r}{-0.14$\,\pm\,$0.06}     &      \multicolumn{1}{l}{\hspace{0.2cm}  292$\,\pm\,$10       }  &   \multicolumn{1}{c}{\hspace{-0.3cm}$-0.11\,\pm\,0.09$}    &   \multicolumn{1}{c}{0.30}            & \multicolumn{1}{r}{\hspace{-0.2cm}$ 0.04\,\pm\,0.02$}  &   \multicolumn{1}{c}{0.55}                 \\
\hline
\end              {tabular*}
\begin{footnotesize}
\begin{flushleft}
{\bf Notes:} \\
$^{a}$ Order of polynomial degree.                                                                                        \\
$^{b}$ Wavelength range used for the SSP fit.                                                                             \\
$^{c}$ Aperture corrected central velocity dispersion.                                                                    \\
$^{d}$ Age- and metallicity gradients as defined in equations \ref{gradientage} and \ref{gradientmetallicity}.            \\
$^{e}$ Coefficient of determination: $R^{2}=1$ $-$ perfect linear relation;  $R^{2}=0$ $-$ no linear relation.            \\
\end{flushleft}
\end{footnotesize}
\end{minipage}
\end              {table*}

\subsection{Line-strength gradients}  
\begin{figure}
\centering
\includegraphics[width=\columnwidth]{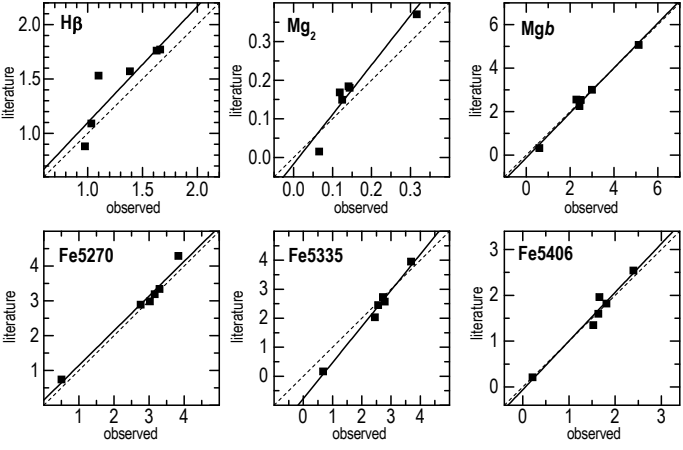}
\caption[Radial  velocity   distribution]{\label{lickcomparison}Comparison   between the measured   lick indices and  the  original lick  data    from
\citet{worthey97} for  the  observed  lick standards. Horizontal axes  present the   measured indices   from  this work while   vertical ones show the
corresponding values  from the literature. Dashed lines indicate   unity while solid ones present  linear least squares fits to the data.}
\end{figure}

Besides  SSP  ages and  metallicities,  the   strongest  absorption  lines  have   been measured  to investigate
line-strength  gradients.  For  that purpose, Lick/IDS   indices as defined in \citet{worthey94}  and   \citet{worthey97}
have been   analysed.   Due   to  the limited  S/N at larger galactic radii,   only  the most prominent   indices  (H$\beta$, Mg$_2$,
Mg$b$, Fe5270, Fe5335, Fe5406) have  been  considered. Figure  \ref{spectra}    highlights    the redshifted  index   bands   in  all  six    ellipticals. The
presented    spectra   have been  smoothed  to   the Lick/IDS resolution.   Other prominent lines    are also marked.  Since the  spectral   resolution of
the Lick/IDS    system amounts to   only $\sim9\,$\AA,  the observed  spectra had to  be degraded in resolution   first. For that purpose, the FWHM of
emission lines  was measured   in the  arc comparison  spectra yielding an estimate of  the actual spectral resolution.  The     spectra   were  then
convolved       with     a       Gaussian    kernel    of     dispersion     ${\sigma     _{{\rm{smooth}}}}     =   {{\rm{[(FWHM}}_{{\rm{Lick}}}^2   -
{\rm{FWHM}}_{{\rm{obs}}{\rm{.}}}^2)/8\ln  2]^{1/2}}$     to match   the  Lick/IDS   system.     The         broadened     spectra         were   subsequently
analysed          with             the            open-source                program             {\tt       indexf}\footnote{version               3.0
$-$          24.08.2005\\ \url{http://www.ucm.es/info/Astrof/software/indexf/indexf.html}}   (\citealt{cardiel07}).     To     estimate
errors  of   the  measured  indices,  again  a   fixed    S/N estimated     at  rest-frame     wavelength      of $\sim5140\,$\AA\space    and   200
Monte    Carlo  simulations  were    considered.  Final  indices and    errors  were then  computed  as    the   mean  of   the    200  Monte
Carlo      simulations.     To   transform  the  observed  indices     to   the  Lick/IDS  system,   Lick standards   have   been   observed. Figure
\ref{lickcomparison}      shows   the   comparison    between     the      measured     indices   and      the     original     Lick  data        from
\citet{worthey97} for   the   observed  standards. Indices   show   only  small  differences      compared to   the   original  Lick   data,   proving
the   reliability   of our measurements.   To   quantify    the deviations  to  the   Lick system,  linear  relations have been  fit  to the
data. To account    for  Doppler   broadening,  indices  were  measured   consecutively in  the  Lick  standard star   spectra  broadened  to 50$-$400
km~s$^{-1}$ in steps of 50  km~s$^{-1}$. Assuming that   $({\rm index}_{\,i,j,\sigma})$  and  $({\rm  index}_{\,i,j,0})$  are the     absorption  line-strengths   of   index  $i$   and  standard   star  $j$   measured  at  velocity  dispersion   $\sigma$  and   0,  then ${\mathcal{R}_{i,\sigma  }}  =
{n^{ -  1}}\sum\nolimits_{j  =  1}^n {{\rm{[inde}}{{\rm{x}}_{i,j,\sigma  }}/} {\rm{inde}}{{\rm{x}}_{i,j,0}}]$   is  the  average  correction    factor
estimated   from all observed   standard   stars  for   index   $i$. Thus,  an   index   in   a    galaxy  with   velocity   dispersion  $\sigma$  was
corrected   by  means  of   $\label{dispcorrect}    {\rm{index}}_{\,i,0}   = {\rm{index}}_{\,i,\sigma  } \cdot \mathcal{R}_{\,i,\sigma  }^{-1}$.   Not
all indices   show   the  same  dependency   with $\sigma$.    While  iron     lines show    a   rather  steep  decline   in index   strengths,
 H$\beta$     and  Mg$_{2}$    show  comparatively     small      dependence.   Correction  functions  $\mathcal{R}_{\,i,\sigma}$   were  estimated by
 averaging  individual correction functions $\mathcal{R}_{\,i,j,\sigma}$ (excluding  obvious outliers)  and     fitting
    third  order    polynomials   to the averaged data.   Ultimately,   all measured indices    were  corrected  for   velocity  dispersion  first  and
subsequently transformed to the published  lick data. Gradients $\nabla  I$ of  individual indices  $I$ were measured  by linear  least  squares
fits of  the form

\begin{equation}
I(r/{r_{{\rm{eff}}}}) = I({r_{{\rm{eff}}}}) + \nabla I\log \left[{r/{r_{{\rm{eff}}}}} \right] \label{lickequation}
\end{equation}

\begin{figure*}
\centering
\includegraphics[width=\textwidth]{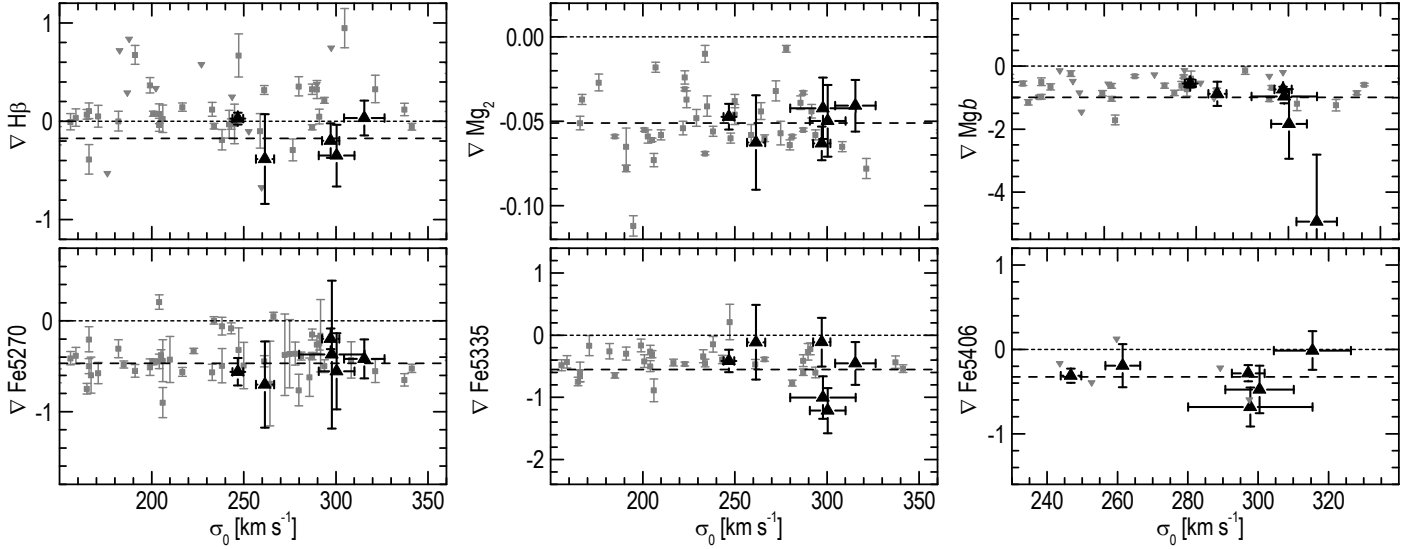}
\caption{\label{gradients2}Line-strength gradients  versus central velocity dispersion. Black  triangles show the observed FCGs while grey  squares
and triangles are index measurements from \citet{kobayashi99} and \citet{rawle}. Dashed lines represent  the average gradient for all  observed fossils
while short-dashed lines indicate zero  gradients. The  H$\beta$ gradient is   not shown for  RX  J0752.7+4557  since  the line is   seen  in emission.  The
very  steep  Mg$b$ gradient of  RX J1548.9+0851 has been excluded for  the estimation of the average Mg$b$ gradient.}
\end{figure*}

so that $\nabla I \equiv \Delta I/\Delta \log \left( {r/{r_{{\rm{eff}}}}} \right)$.  As for ages and metallicites, indices at a  certain radius   were
averaged from both sides of  the galaxy centre. Obvious  outliers  at larger  radii deviant from   the overall gradient  trend  were  excluded for the
fit.  Figure   \ref{gradients2} shows   the  resulting   gradient slopes   $\nabla I$   with respect   to  the   central velocity  dispersions of each
galaxy. Dashed lines show the   avereage line-strength gradient for all   galaxies while short-dashed lines highlight  flat gradients for  comparison.
In  the case  of  Mg$b$, the   very   steep  gradient  for  RX  J1548.9+0851  has been    excluded for    the estimation   of the    average  gradient
slope. We compare our index  gradients with those  from \citet{kobayashi99}  and \citet{rawle} and find similar   gradient slopes. Only for Mg$b$  and
Fe5335  we  find gradients  that are  much  steeper  than in  our comparison   sample. Table    \ref{ellipticalsabundancegradients} lists   the  line-strength
 gradients for   each index   and also   gives  the  coefficient  of  determination for each    fit. All   metal  sensitive   lines (Fe    and
Mg)   show  negative  gradients  confirming   the    negative  metallicity    gradients   derived    from  the   SSP    fits.     Only     Fe5406   in
RX1548.9+0851  exhibits  a flat gradient. The   age  sensitive    H$\beta$ index   exhibits  both  negative and  slightly  positive   gradients.   The
steep  positive  gradient  for H$\beta$  in RX     J0752.7+4557   results     from  the   line being     observed  in emission.  Hence,    this  slope
is not    shown    in   Fig.\ \ref{gradients2}.  The    observed   line-strength   gradients  follow   the    trend derived   from   the   SSP   fits,
clearly   indicating negative metallicity gradients.     To  compare   our  index   gradients     with  the  metallicity  gradients     from
ULySS,  we   consider   index-metallicity  relations  from    \citet{wortheymodels94}  SSP models. In particular,   we fit linear relations  to the
\citet{wortheymodels94} indices Mg$_2$, Mg$b$,     Fe5270, Fe5335,  Fe5406 for metallicities  $-0.5  \le  [{\rm{Fe/H}}]  \le 0.5$  and  galaxies  with
12Gyr.   These relations are  of the  form    $[{\rm{Fe/H}}] =   p  \cdot {\rm{index}}  -  d$  so  that  an index  gradient  $\nabla  I$   corresponds
to  a   metallicity   gradient     $\nabla_{[\rm{Fe/H}]}=p  \cdot  \nabla   I$.    We  derive  values  of    $p_{\,\rm{Mg}_{2}}=5.48$,  $p_{\,{\rm
Mg}b}=0.44$,  $p_{\,{\rm   Fe5270}}=0.63$, $p_{\,{\rm  Fe5335}}=0.52$,  and     $p_{\,{\rm  Fe5406}}=0.79$. Applying   these conversion   factors   to
the average  line-strength gradient  slopes illustrated in    Fig.\ \ref{gradients2},  we  find  average metallicity  gradients   of    $\nabla_{[{\rm
Fe/H}]}=-   0.28\pm0.05$      ($\rm{Mg_{2}}$),  $\nabla_{[{\rm      Fe/H}]}=-0.44\pm0.22$     ($\rm{Mg}b$),     $\nabla_{[{\rm   Fe/H}]}=-0.29\pm0.11$
($\rm{Fe5270}$), $\nabla_{[{\rm Fe/H}]}=-0.29\pm0.24$ ($\rm{Fe5335}$), $\nabla_{[{\rm Fe/H}]}=-0.26\pm0.18$ ($\rm{Fe5406}$)  for the observed fossils.
Except for  the systematically lower value    derived from  Mg$b$,   all  gradient slopes     show very  similar    values  resulting in    an average
metallicity    gradient   of   $\nabla_{[{\rm    Fe/H}]}=-  0.31\pm0.07$.   This  result   is   consistent    with   the    average   gradient   slope
$\nabla_{[\rm{Fe/H}]}=-0.19\pm  0.08$  derived from   full-spectrum fitting   within the    error bars.   Individual  gradient    slopes  are     also
consistent  with   the  metallicity    gradients  derived   from  full-spectrum   fitting within  the  corresponding  error bars  except for NGC1132,
showing a systematically flatter gradient. We derive $\nabla_{[\rm{Fe/H}], \, NGC1132}=-0.26\pm  0.05$, $\nabla_{[\rm{Fe/H}], \, RXJ0752}=-0.39\pm  0.15$,
$\nabla_{[\rm{Fe/H}], \, RXJ0844}=-0.22\pm  0.12$, $\nabla_{[\rm{Fe/H}], \, RXJ1152}=-0.28\pm  0.16$, $\nabla_{[\rm{Fe/H}], \, RXJ1520}=-0.41\pm  0.16$  and
$\nabla_{[\rm{Fe/H}], \, RXJ1548}=-0.18\pm  0.12$  from the  averages of  all line-strength  gradients.  However,    we focus  our results on the  full-
spectrum fitting technique  since every pixel is  considered  for the  determination of  metallicity   instead  of only a few  individual indices.

\begin{table*}
\begin{minipage}[t]{520pt}
\caption{\label{ellipticalsabundancegradients} Line-strength gradients as defined in equation \ref{lickequation}.}
\centering
\renewcommand{\footnoterule}{}
\begin            {tabular*}{\textwidth}{@{\extracolsep{\fill}}p{2.5cm}p{0.3cm}p{1.8cm}p{1.3cm}p{1.6cm}p{0.6cm}p{0.3cm}p{0.3cm}p{1cm}p{0.3cm}p{0.3cm}p{0.3cm}p{0.3cm}}
\hline
\hline
 \multicolumn{1}{c}{galaxy}         &    \multicolumn{1}{c}{$\nabla \rm{\, H}\beta$}    &  \multicolumn{1}{c}{$R^{2\,a}$}   & \multicolumn{1}{c}{$\nabla \rm{\, Fe5270}$}    &      \multicolumn{1}{c}{$R^{2\,a}$}  &        \multicolumn{1}{c}{$\nabla \rm{\, Fe5335}$}   &         \multicolumn{1}{c}{$R^{2\,a}$}    &   \multicolumn{1}{c}{$\nabla \rm{\, Fe5406}$}     &           \multicolumn{1}{c}{$R^{2\,a}$}    &                 \multicolumn{1}{c}{$\nabla \rm{\, Mg_{2}}$}     &  \multicolumn{1}{c}{$R^{2\,a}$}       & \multicolumn{1}{c}{$\nabla \rm{\, Mg}b$}    &  \multicolumn{1}{c}{$R^{2\,a}$}            \\
\hline
NGC 1132                            &    \multicolumn{1}{r}{ 0.0$\,\pm\,$0.1}           &   \multicolumn{1}{c}{0.02}        &  \multicolumn{1}{c}{-0.6$\,\pm\,$0.2}          &       \multicolumn{1}{c}{0.56}       &        \multicolumn{1}{c}{-0.4$\,\pm\,$0.2}          &       \multicolumn{1}{c}{0.32}            &  \multicolumn{1}{r}{-0.3$\,\pm\,$0.1}             &    \multicolumn{1}{c}{0.56}                 &                   \multicolumn{1}{c}{-0.05$\,\pm\,$0.01}        &   \multicolumn{1}{c}{0.78}            & \multicolumn{1}{c}{-0.6$\,\pm\,$0.1}        &   \multicolumn{1}{c}{0.61}                 \\
RX J0752.7+4557                     &    \multicolumn{1}{r}{ 1.8$\,\pm\,$0.1}           &   \multicolumn{1}{c}{0.99}        &  \multicolumn{1}{c}{-0.4$\,\pm\,$0.8}          &       \multicolumn{1}{c}{0.09}       &        \multicolumn{1}{c}{-1.0$\,\pm\,$0.3}          &       \multicolumn{1}{c}{0.81}            &  \multicolumn{1}{r}{-0.7$\,\pm\,$0.2}             &    \multicolumn{1}{c}{0.81}                 &                   \multicolumn{1}{c}{-0.04$\,\pm\,$0.02}        &   \multicolumn{1}{c}{0.73}            & \multicolumn{1}{c}{-1.0$\,\pm\,$0.2}        &   \multicolumn{1}{c}{0.91}                 \\
RX J0844.9+4258                     &    \multicolumn{1}{r}{-0.2$\,\pm\,$0.2}           &   \multicolumn{1}{c}{0.24}        &  \multicolumn{1}{c}{-0.2$\,\pm\,$0.1}          &       \multicolumn{1}{c}{0.43}       &        \multicolumn{1}{c}{-0.1$\,\pm\,$0.4}          &       \multicolumn{1}{c}{0.02}            &  \multicolumn{1}{r}{-0.3$\,\pm\,$0.1}             &    \multicolumn{1}{c}{0.69}                 &                   \multicolumn{1}{c}{-0.06$\,\pm\,$0.01}        &   \multicolumn{1}{c}{0.91}            & \multicolumn{1}{c}{-0.7$\,\pm\,$0.1}        &   \multicolumn{1}{c}{0.95}                 \\
RX J1152.6+0328                     &    \multicolumn{1}{r}{-0.4$\,\pm\,$0.5}           &   \multicolumn{1}{c}{0.19}        &  \multicolumn{1}{c}{-0.7$\,\pm\,$0.5}          &       \multicolumn{1}{c}{0.42}       &        \multicolumn{1}{c}{-0.1$\,\pm\,$0.6}          &       \multicolumn{1}{c}{0.01}            &  \multicolumn{1}{r}{-0.2$\,\pm\,$0.3}             &    \multicolumn{1}{c}{0.16}                 &                   \multicolumn{1}{c}{-0.06$\,\pm\,$0.03}        &   \multicolumn{1}{c}{0.72}            & \multicolumn{1}{c}{-0.9$\,\pm\,$0.4}        &   \multicolumn{1}{c}{0.72}                 \\
RX J1520.9+4840                     &    \multicolumn{1}{r}{-0.3$\,\pm\,$0.3}           &   \multicolumn{1}{c}{0.24}        &  \multicolumn{1}{c}{-0.6$\,\pm\,$0.4}          &       \multicolumn{1}{c}{0.30}       &        \multicolumn{1}{c}{-1.2$\,\pm\,$0.4}          &       \multicolumn{1}{c}{0.74}            &  \multicolumn{1}{r}{-0.5$\,\pm\,$0.3}             &    \multicolumn{1}{c}{0.41}                 &                   \multicolumn{1}{c}{-0.05$\,\pm\,$0.02}        &   \multicolumn{1}{c}{0.65}            & \multicolumn{1}{c}{-1.8$\,\pm\,$1.1}        &   \multicolumn{1}{c}{0.41}                 \\
RX J1548.9+0851                     &    \multicolumn{1}{r}{ 0.0$\,\pm\,$0.2}           &   \multicolumn{1}{c}{0.01}        &  \multicolumn{1}{c}{-0.4$\,\pm\,$0.2}          &       \multicolumn{1}{c}{0.49}       &        \multicolumn{1}{c}{-0.5$\,\pm\,$0.3}          &       \multicolumn{1}{c}{0.30}            &  \multicolumn{1}{r}{ 0.0$\,\pm\,$0.2}             &    \multicolumn{1}{c}{0.00}                 &                   \multicolumn{1}{c}{-0.04$\,\pm\,$0.02}        &   \multicolumn{1}{c}{0.64}            & \multicolumn{1}{c}{-4.9$\,\pm\,$2.1}        &   \multicolumn{1}{c}{0.57}                 \\
\hline
\end              {tabular*}
\begin{footnotesize}
\begin{flushleft}
{\bf Notes:} \\
$^{a}$ Coefficient of determination: $R^{2}=1$ $-$ perfect linear relation;  $R^{2}=0$ $-$ no linear relation.            \\
\end{flushleft}
\end{footnotesize}
\end{minipage}
\end              {table*}

\subsection{Velocity dispersions}
\label{dispersions}
To estimate the   dynamical mass  of  the observed   ellipticals,   line  of  sight   velocity  dispersions (LOSVDs) have  been derived   with  ULySS.
By   fitting SSP  models    to  the observed  spectra, the  package   also  measures  the  broadening      of  each SSP  model    to  match  the
observed    spectrum,   allowing for     a   proper determination of    LOSVDs. ULySS determines    the     broadening  relative to  the     SSP  model
dispersion    $\sigma_{\rm{model}}$,    amounting    to  13   km   s$^{-1}$   for  the  Pegase HR  models.  In  addition,  the
instrumental   dispersion  $\sigma   _{\rm{instr}}$   of     the   spectrograph   was   taken  into  account  by   measuring  FWHMs   of   emission
lines  in the arc    lamp comparison spectra  taken  during the  night  in  the   same setup  as     the  galaxies. Based  on   the average of several
lines a final value  of  $\sigma    _{\rm{instr}}=95$ km  s$^{-1}$    was considered\footnote{$\rm{FWHM}  = 2\sqrt  {2\ln  2}   \cdot  \sigma$}. Final
LOSVDs   were  ultimately  computed   via   the   relation:  $\sigma  _{\rm{phys}}^2  =  \sigma _{\rm{ulyss}}^2   +  \sigma  _{\rm{model}}^2  - \sigma
_{\rm{instr}}^2$.  The observed  central velocity dispersions   have been measured    in the inner  seeing-limited  bin of  each galaxy.  This central
aperture corresponds to different physical radii   based  on the distance  of each galaxy.  To account  for
this inconsistency, we  applied an  aperture  correction, transforming  the  observed dispersions to  a system that  is
independent of  distance.   For    that    purpose,   the   prescription          ${\sigma   _{{\rm{0,corr}}}}    =   {\sigma   _{{\rm{obs}}}}   \cdot
{[   {8({{r_{{\rm{aperture}}}}/{r_{{\rm{e}}}}} )} ]^{0.04}}$  has been applied, were $r_{\rm{aperture}}$ is the central aperture and $r_{\rm{e}}$  the
effective radius (\citealt{jorgensen,wegner}). Table  \ref{ellipticalsresults}  shows the resulting, aperture corrected central velocity  dispersions  ${\sigma   _{{\rm{0,corr}}}}$. 
  The  integrated  velocity  dispersions   $\sigma$  were  used  to  correct  the   measured  absorption  indices  for   Doppler
broadening.   The  studied FCGs exhibit an average central velocity dispersion of $271\,\pm\,28$  km s$^{-1}$. Comparing  this  value  with the  LOSVDs  of  $\sim9000$
ellipticals from \citet{bernardirelations} shows   that the investigated galaxies  are among the  most-massive ellipticals in the universe.

\subsection{The  fundamental  plane}
Elliptical   galaxies   show  a    tight   relation  between   their    effective  radius   $r_{e}$,   the   mean
effective   surface   brightness     $\left\langle\mu\right\rangle_{e}=-2.5\log  \left\langle I   \right\rangle_{e}$,  and   the   central    velocity
dispersion   $\sigma   _0$    known  as    the  fundamental   plane    (FP)   (\citealt{djorgovskidavis87,dressler87}):   $r_e  \propto  \sigma   _0^A
\left\langle {I_e} \right\rangle   ^B$, with varying  values of  $A$  and  $B$ by   different authors  (see \citealt{magoulas,bernardi03}).   Since   this
fundamental   plane  spans about     three  orders    of magnitude     in luminosity,   exhibiting    very   low    residual   scatter    and   little
variation     with  environment,  it   suggests  a  uniform  formation process  of   ellipticals.  The  FP  can  be  reconstructed by  means  of   the
virial theorem which implies  values  of   $A=2$ and $B=-1$   and  a    constant   mass-to-light ratio along the    plane. However,   observed  values
of  $A=1.49$   and $B=-0.75$   from a    sample  of  $\sim$ 9000   SDSS early-type  galaxies  by  \citet{bernardi03} indicate   a  tilted  plane  that
deviates from  the virial   prediction.

To     study    the        location     of     the       observed      FCGs     on        the    FP      we        modeled     the    SDSSIII      DR9
surface  brightness      profiles      with  {\texttt   GALFIT}\footnote{\url{http://users.obs.carnegiescience.edu/peng/work/galfit/galfit.html}}   by
fitting a     \citet{devaucouleurs48}    $r^{1/4}$       law   to    the galaxy   surface brightness   distribution.   Nearby   stars   were  used  to
estimate the   PSF   and  correct     for    seeing.    By   applying    a   \citet{devaucouleurs48}    $r^{1/4}$  law,   the         mean   effective
surface brightness      $\left\langle\mu\right\rangle_{e}$     can      be     computed       as      $\left\langle\mu\right\rangle_{e}=\mu_{e}-1.393$
(see \citealt{grahamdriver05}). The  measured    surface  brightness  was    corrected    for $(1+z)^4$  cosmological  dimming,   extinction   and $K-
$correction,   taken from   the SDSS.  Effective  radii were  corrected   for ellipticity  so that $r_e=r_{e\rm{,obs}}   \sqrt{b/a}$,   where  $b$  and
$a$   are the semi$-$minor and semi$-$major    axes, respectively. As mentioned in Sect.\  \ref{dispersions} we applied an aperture correction  to our
central velocity dispersions. The same procedure was applied in \citet{bernardi03a} allowing us to directly compare our measurements with their work.

\begin{figure}
\centering
\includegraphics[width=\columnwidth]{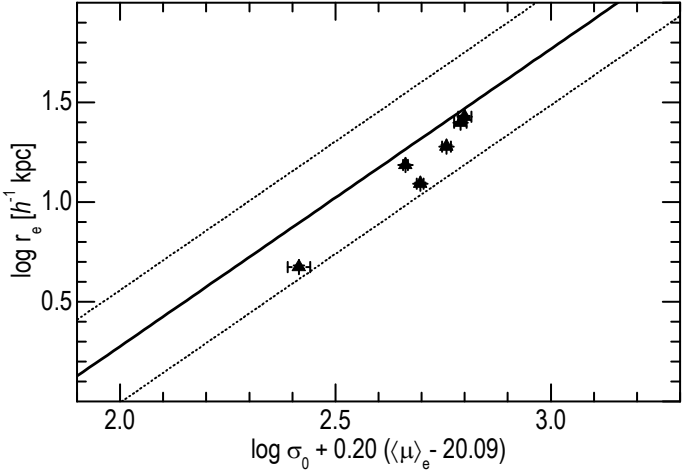}
\caption{\label{plane} The observed FCGs on the fundamental plane. The  solid line shows the best-fit FP from  \citet{bernardi03} obtained  from $\sim
9000$  early-type galaxies in the SDSS in the  redshift  range $0.01 \le z \le 0.3$ while   the dotted lines  present the  3 $\sigma$ deviations  from
the   fit. Error  bars  for effective  radii and  surface  brightness  were determined   with GALFIT  while velocity  dispersion errors were estimated
with ULySS.}
\end{figure}

Figure \ref{plane} shows the  observed FCGs on the  fundamental  plane from \citet{bernardi03},   obtained from $\sim  9000$  early-type galaxies   in
the  SDSS  in the redshift  range $0.01  \le   z \le  0.3$ (solid  line). The   dotted  lines   present   3 $\sigma$  deviations   from  the    fit.
All FCGs  are  located within these  thresholds, following the   SDSS fundamental plane. This result supports the  idea that these objects are relaxed
systems that had  enough time to   virialise. However, all  galaxies are found   below the FP,  indicating that they  are either more  massive or more
compact compared to the majority of ellipticals. Given that these galaxies  are expected to have accumulated enough mass from $L^{*}$ galaxies of  the
progenitor group, it is plausible that the observed fossils show systematically higher masses compared to average ellipticals.

\section{Discussion}  Radial     gradients     of   age,        metallicity,   and  individual   absorption    line-strengths           have      been
extensively                studied      in                  the                      past                  for                                 several
elliptical            galaxies (\citealt{davies93,fisher95,sanchezblazquez,fatmareda,annibali07,baes,rawle,loubser},     and    references   therein).
These  gradients   are   in general  well    fit  by   power-laws   and    their   slopes    provide     useful  constraints      on    the  formation
scenario  of  ellipticals. Following     this   approach,  we    have  observed  six  fossil   central    ellipticals  (FCGs)    spectroscopically  to
study   their   stellar populations  and   measure    age, metallicity, and line-strength   gradients of the  strongest absorption   features  to shed
light   on the   formation and  evolution of fossil systems. For   that  purpose we obtained   medium-resolution   long-slit spectra at  the WHT   and
made use of  the full-spectrum fitting package ULySS to fit SSP models to  our data.

\begin{figure*}
\centering
\includegraphics[width=\textwidth]{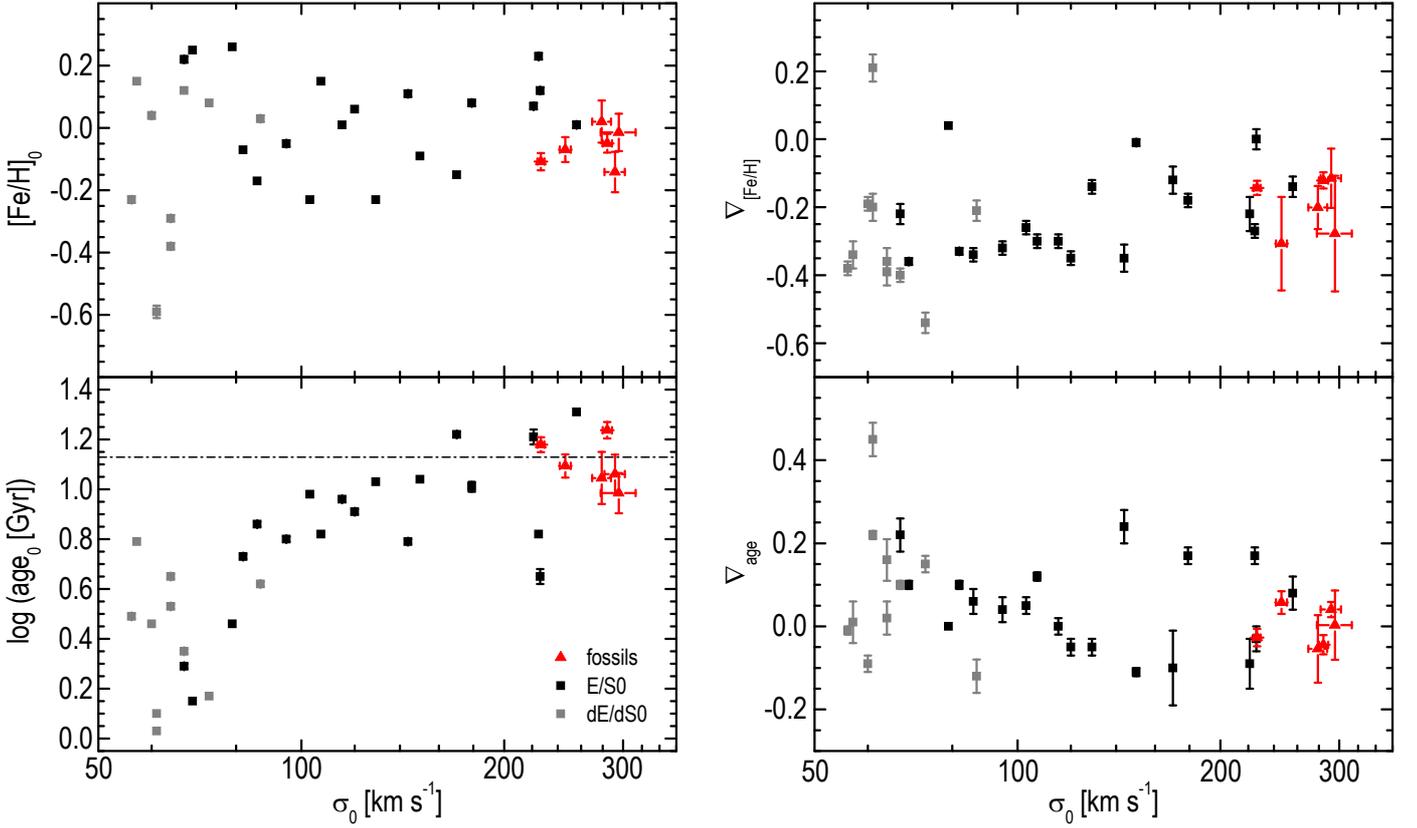}
\caption{\label{gradientssigma} Left  panel: Central  age and  metallicity versus  central velocity  dispersion. Right  panel: Gradients  of age   and
metallicity  versus central   velocity dispersion. Black and  grey  squares are early-type galaxies (E/S0, dE/dS0)   from  \citet{koleva11} while  red
triangles show  the  investigated FCGs.  Error bars for  age,  metallicity and  velocity dispersion were  derived from the  ULySS fits while  gradient
errors were computed from the corresponding linear-least squares fits.}
\end{figure*}

\begin{figure}
\centering
\includegraphics[width=\columnwidth]{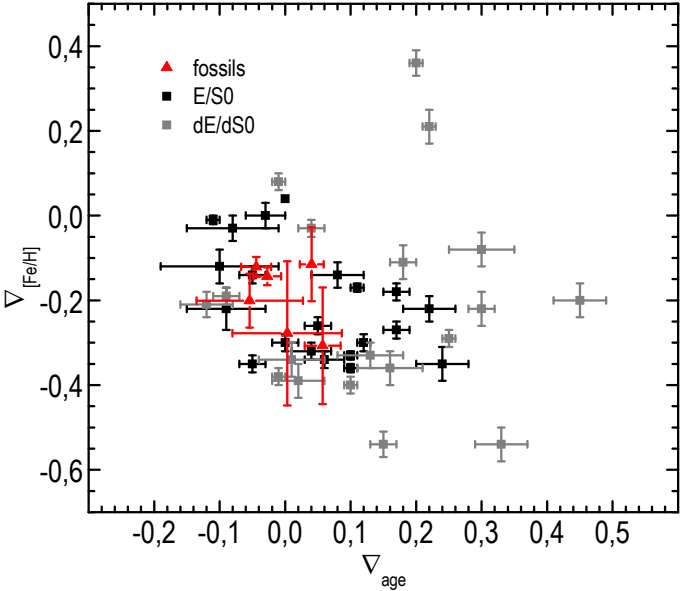}
\caption{\label{gradientsgradients} Gradients of metallicity versus gradients of age. Symbols as in Fig.\ \ref{gradientssigma}.}
\end{figure}

Figure \ref{gradientssigma} shows central velocity  dispersions, ages, and metallicities  of  our FCGs (red triangles) in  comparison with the  sample
of   early-type  galaxies  presented  in  \citet{koleva11}.   Black  squares  are  E  and   S0   galaxies  in  the  Virgo  and   Fornax Clusters  (see
also  \citealt{spolaor,bedregal,vazdekis}) while  grey symbols    show early-type  dwarfs from    \citet{derijcke}. The  sample of    \citet{koleva11}
provides an ideal   comparison  for our  measurements since  it  covers a broad   range   in galaxy  mass  and  stellar  populations are  analysed  as
highlighted   in   this  work,   i.e.   all   galaxies  are    fit  with   Pegase  HR    models   within   ULySS.   Exhibiting    an average   central
velocity  dispersion of $\sigma_0=271\pm28$   km  s$^{-1}$,   our  sample    comprises some   of  the  most-massive  ellipticals    in  the  universe,
populating   the high  end mass-regime of early-type  galaxies. This is   not  surprising given the fact  that all  observed ellipticals are  expected
to be  the remnants   of  the  merger of  multiple $L^{*}$  galaxies.  The   mass-metallicity relation  is evident in Fig.\ \ref{gradientssigma}, more
massive galaxies   retaining their  metals  much   easier than   less-massive ones.    We  find  similar central   metallicities around  [Fe/H]$_{0}=-
0.06\pm0.06$ for  our fossils  compared to  the  cluster  ellipticals in  Virgo   and  Fornax of  similar  mass. Central ages are  old and exhibit  an
average   value of $12.8   \pm 0.8$ Gyr.   Except for  one  galaxy,  all observed   ellipticals exhibit  SSP  ages  older than     10  Gyr  indicating
formation   epochs  at   $z >  2$. RX  J0752.7+4557 shows  an  intermediate  age of   $8.0\pm 1.0$  Gyr   suggesting  a more extended   star formation
history   to   $z=1$.  This  galaxy    is  also   the  only  one    in   our   sample  with    prominent  nebular    emission   lines   (see    Fig.\
\ref{spectra}) and  it was  classified as  AGN by  multiple authors  (\citealt{hess,veroncetty}). Metallicity  gradients in  our sample  are found  to
be throughout     negative  with   an  average    slope of   $\nabla_{[\rm{Fe/H}]}=-0.19\pm  0.08$  while    age    gradients are   flat,   scattering
around $\nabla_{\rm{age}}=0.00\pm   0.05$. Figures\    \ref{gradientssigma} and   \ref{gradientsgradients} show   that the     observed gradients  are
similar  to those   of  our  control  sample,  suggesting that   the physical  processes   involved in   the  formation   of fossils    do not  differ
systematically  from those  forming cluster  ellipticals.  The full-spectrum fitting technique applied here   is a powerful tool in breaking  the 
age-metallicity degeneracy of old  stellar populations. However, to confirm our SSP fitting results, we  also measured radial  gradients for    individual
line-strengths. Metal-sensitive  aborption  features  such  as   Mg$_2$, Mg$b$,   Fe5270,  Fe5335 all  show  negative radial gradients while  the 
age-sensitive H$\beta$ index shows   both positive  and negative   gradient slopes, scattering around    a flat gradient. We   compare our index gradients
with those  from \citet{kobayashi99} and \citet{rawle} and find similar  gradient  slopes. Only for Mg$b$ and Fe5335  we find gradients that  are much
steeper than in our  comparison  sample.  From our  index  measurements we derive  an average  metallicity   gradient   of  $\nabla_{[{\rm   Fe/H}]}=-
0.31\pm0.07$.  This   result  is  lower but  consistent   with   the   average   gradient    slope  $\nabla_{[\rm{Fe/H}]}=-0.19\pm  0.08$ derived from
full-spectrum fitting  within the     error bars.    Individual  gradient     slopes  are      also consistent   with   the   metallicity    gradients
derived   from  full-spectrum   fitting within  the  corresponding  error bars  except for NGC1132, showing a systematically flatter gradient.

Two competing formation  scenarios have been  proposed  for the  formation of elliptical  galaxies. Either  ellipticals  are formed monolithically  by
dissipational collapse  of a pristine gas cloud (\citealt{larson74,carlberg84}) or hierarchically through successive mergers of disk galaxies or  many
dwarf galaxies  (\citealt{kauffmannwhite}). In   the  dissipative  collapse      model,   gas  gets  continuosly  enriched  in  the  course  of 
 star-formation  as  it  flows  towards  the  galaxy centre,  resulting in   steep metallicity   gradients. In   this scenario   gradient slopes   have been
predicted  to   be  as  steep  as   $\nabla_{Z}=\Delta \log Z/\Delta \log r =-0.35$  (\citealt  {larson74}), $\nabla_{Z}=-1.0$   (\citealt  {larson75}),  and $\nabla_{Z}=-0.5$  (\citealt
{carlberg84}).   Indeed,        negative  metallicity  gradients,  observed  as  radial  gradients of  absorption  line-strengths, are       a  common
feature  in   ellipticals,  however   exhibiting typically  shallower slopes  around $\nabla_{Z}  =-0.3$. Monolithic  collapse also
predicts a strong correlation between galaxy mass and metallicity gradients since deeper potential wells retain more metals.

In the framework of  hierarchical mass assembly, ellipticals  are the  results  of  galaxy mergers. \citet{white80} has   shown  that mergers  flatten
and erase the metallicity     gradients  of the progenitor  galaxies  due   to   orbital   mixing of   stars. \citet{mihoshernquist94}   modelled  the
merger of   disk    galaxies and    concluded    that   pure  stellar    disks  do   not     reproduce    the power-law  gradients    observed      in
ellipticals unless      a   metal-rich starburst is      involved.  \citet   {kobayashi04} studied   the evolution   of metallicity  gradients in  the
framework  of $\Lambda$CDM cosmology,   in which the formation   and  evolution   of ellipticals   is expected   to be   driven by  mergers.  Allowing
for   various  merging  histories  for    ellipticals,  ranging  from   quasi-monolithical    collapse    to    multiple   major     mergers,   \citet
{kobayashi04} found   no    correlation  between gradient   slopes and  galaxy  mass  and stated  that  the  variety of   gradients originates in  the
different  merging histories of ellipticals in  the  sense that  galaxies that form   monolithically have steeper gradients than  galaxies   resulting
from major  mergers.  The change  in  gradient   slope  depends   strongly    on   the   mass   ratio   of   the merging   galaxies  and   the induced
star    formation     during   the     merger.  Typical     gradients  were found to be       $\nabla_{Z}=-0.22$, $\nabla_{[\rm{Fe/H}]}=-0.38$  for    major     mergers
and $\nabla_{Z}=-0.30$, $\nabla_{[\rm{Fe/H}]}=-0.45$  for   non-major mergers.  Simulating  gas-rich  disk galaxy mergers,  \citet{hopkinsdissipation} also found that  the
amount of  dissipation, characterizing  the merger-induced starbursts involved    during  the  merger,   is   the  essential  factor   in  determining
the final  gradient slopes. \citet{dimatteo09} investigated    the shape      of   metallicity    gradients      resulting     from    dissipationless, (gas-free)
dry mergers with   the  conclusion that the   gradients    strongly depend   on    the  gradient   shapes    of      the     progenitor
galaxies.    Two galaxies     with   similar gradients  will result  in  a flatter      gradient  in the  final   galaxy.   However,    the    initial
gradient      can     also   be   preserved    or even   increased     by     the   merger      if   the   slope   of the   companion     galaxy    is
sufficiently  steep.

\begin{figure}
\centering
\includegraphics[width=\columnwidth]{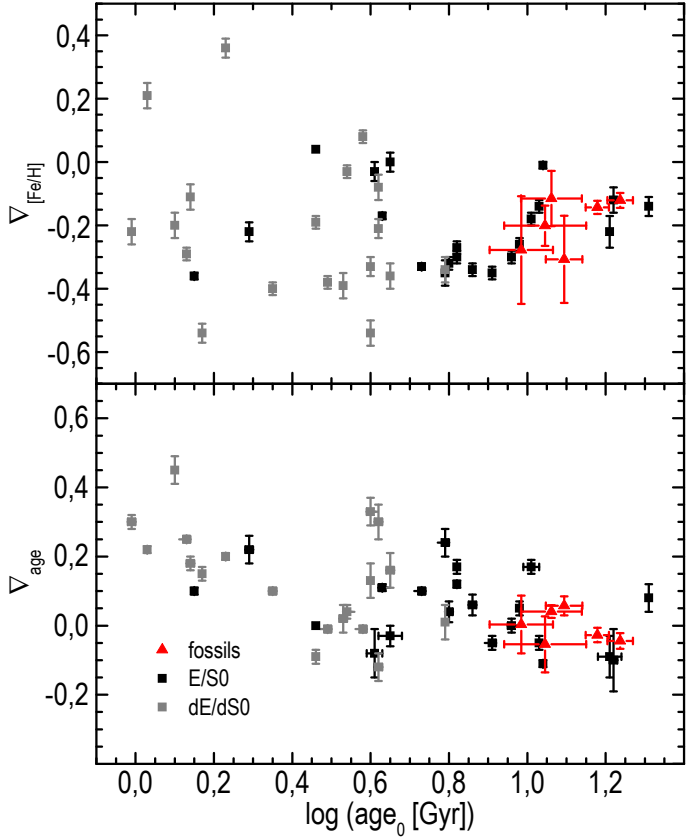}
\caption{\label{gradientsvsage} Gradients of age and metallicity versus central age. Symbols as in Fig.\ \ref{gradientssigma}.}
\end{figure}

The    comparatively    shallow   metallicity    gradients    of    our   sample ($\nabla_{[\rm{Fe/H}]}=-0.19\pm 0.08$) with respect  to the predicted
gradients from     monolithic  collapse  for the  most-massive  galaxies  ($\nabla_{Z}=-0.5$; \citealt  {carlberg84})  suggest   that   major  mergers
have    been  involved   in   the    formation of   fossils,  continuosly   flattening initial   gradients   to     the    present     day    values.
Especially when  comparing our results with  the simulated gradients in \citet  {kobayashi04}, our gradients derived from both ULySS and lick  indices
match the distribution from  major mergers whereas non-major  mergers show much steeper  gradients. \citet {kobayashi04}  argued  that the   spread in
metallicity  gradients and hence   the formation of  ellipticals  cannot be explained  by   major mergers  or monolithical   collapse  alone  but both
processes might  be involved  in shaping the final gradients in ellipticals. However,  if fossils    originate from   an atypical, top-heavy  luminosity function -- the failed group scenario
 \citep{mulchaeyzabludoff},  with  the magnitude  gap being in  place  a  priori, i.e.\  the   central elliptical formed
monolithically,  no  $L^{*}$  galaxies would  have  been   present to allow for  subsequent  major  mergers after  an initial monolithic collapse  and
we would expect to observe much steeper gradients. Hence  we conclude  that fossils are not failed groups   
 but indeed formed through successive major mergers  of $L^{*}$ galaxies.

\citet{spolaor}  also found that massive galaxies with $\log \sigma_{0}>2.15$  show in general flatter gradients than low-mass galaxies, but exhibit a
much larger scatter in gradient slopes. The flatter gradients were also  attributed to galaxy mergers which should be the driving process in  building
up galaxies above this  mass   limit. The   large  scatter in  gradient  strengths   was  interpreted  to  be   the  consequence  of  various  merging
histories  with different amounts of gas involved  during the  mergers. The  fossil gradients in this  work  do not show a  large spread  in  gradient
slopes suggesting similar merging histories. Figures \ \ref{gradientsgradients} and \ref{gradientsvsage}   show trends between  gradient strength  and
central  age for  the Virgo  and  Fornax early-type galaxies.  There   also   seems to  be a   slight  trend between age   and  metallicity gradients,
steeper   negative metallicity gradients showing more positve  age gradients. All observed fossils  fall  on these relations. However,  because of the
low-number statistics of  our sample it  is impossible to  conclude if fossils  systematically follow these trends.
We also compared our fossils to  the FP of  \citet{bernardi03} and conclude   that our  sample   follows  their   FP    within   $3\sigma$  confidence
intervals, indicating   that they are   dynamically    relaxed  systems, at least in their central parts \citep{mendezabreu12}. Interestingly,  all our galaxies  fall below the  FP, suggesting that  they are either too
compact for  given surface  brightness and  velocity dispersion  or that  they are  too massive   for a  given effective radius. The latter seems very
plausible since these galaxies are expected to constitute  the mass of several $L^{*}$ galaxies.

\section{Conclusion}
We observed  spatially resolved  stellar population  parameters in   a sample  of fossil  ellipticals. Exhibiting     an   average  central   velocity
dispersion  of  $\sigma_0=271\pm28$  km  s$^{-1}$, our   sample   comprises   some   of   the  most   massive   galaxies  in   the  universe, expected
for  the  remnants  of  multiple  mergers  of  $L^{*}$  galaxies.  We  find  comparatively   flat  ($\nabla_{[\rm{Fe/H}]}=-0.19\pm 0.08$)  metallicity
gradients  that   suggest  a hierarchical   merger scenario  for  fossils.    If fossils  were failed groups,  arising  from an   atypical  luminosity
function  where the  $2^{\rm{mag}}$ gap was    in place  a  priori,   then FCGs  would  have  formed  monolithically, not  experiencing   any major
mergers  during   their  evolution     due     to      the     absence     of    $L^{*}$     galaxies.      Models    of     monolithic     collapse
predict   steep     metallicity gradients  ($\nabla_{Z}=-0.5$) in  such a  formation scenario   especially for  massive galaxies. Such steep
gradients  are not observed in this work.   Hence   we conclude that the observed fossils  indeed formed through  major mergers of  $L^{*}$  galaxies.
The low scatter  of gradient slopes  suggests a similar   merging history for   all  galaxies  in  our  sample.   Our   observations   clearly  suffer
from   low-number   statistics and   a   much    larger  sample   of   FCGs    will  be  needed  to  systematically quantify   spatially   resolved
stellar population  parameters   in these galaxies.   Measurements  of  a   larger  sample  will     then help    to   probe  the uniformity  of the
formation  of these systems.

\begin{acknowledgements}
This work is based on observations made with the William  Herschel Telescope (WHT) operated on the island of La Palma by  the Isaac Newton Group (ING)
in  the  Spanish Observatorio  del  Roque  de los  Muchachos of   the Instituto  de Astrof\'{i}sica  de  Canarias. We have made   use    of        the
astronomical       data   reduction     software      {\texttt  {IRAF}}     which     is    distributed      by   the   National   Optical   Astronomy
Observatories,      which  are      operated   by     the  Association    of Universities    for  Research    in  Astronomy  Inc.,  under  cooperative
agreement  with  the National  Science  Foundation.  The publication  was supported  by the  Austrian Science  Fund (FWF).  PE  was  supported  by the
University of  Vienna  in the frame of the  Initiative Kolleg (IK) \textit{The Cosmic Matter  Circuit} I033-N and acknowledges  support from  FONDECYT
through  grant 3130485. We thank  Mina Koleva, Robert Proctor, Claudia Mendes de Oliveira, Christoph Saulder and Bodo Ziegler for useful  discussions.
We  also thank the anonymous referee who read the manuscript carefully and  helped to improve the presentation of the paper.
\end{acknowledgements}

\end{document}